\documentstyle[prd,aps]{revtex}

\def\ut#1{\rlap{\lower1ex\hbox{$\sim$}}#1{}}
\begin{document}

\title{The Constraint Algebra of Quantum Gravity \\
in the Loop Representation}

\author{Rodolfo Gambini}
\address{
Instituto de F\'{\i}sica, Facultad de Ingenier\'{\i}a, \\
J. Herrera y Reissig 565, C. C. 30  Montevideo, Uruguay}

\author{Alcides Garat}
\address{
Instituto de F\'{\i}sica, Facultad de Ciencias, \\
Tristan Narvaja 1674, Montevideo, Uruguay}

\author{Jorge Pullin}
\address{
Center for Gravitational Physics and Geometry\\
Department of Physics\\
The Pennsylvania State University\\
University Park, PA 16802}

\date{April 29 1993}

\maketitle

\begin{abstract}

We study the algebra of constraints of quantum gravity in the loop
representation based on Ashtekar's new variables. We show by direct
computation that the quantum commutator algebra reproduces the
classical Poisson bracket one, in the limit in which regulators are
removed. The calculation illustrates the use of several
computational techniques for the loop representation.

\end{abstract}

\pacs{04.60.+n,02.40+m}
\vspace{-11.5cm}
\begin{flushright}
\baselineskip=15pt
CGPG-94/4-3  \\
gr-qc/9404059\\
\end{flushright}
\vspace{10cm}

\section{Introduction}

An important element in the classical formulation of a canonical
theory which has constraints is their algebra. If taking Poisson
brackets of the constraints of the theory leads to quantities that can
be expressed as combinations of the constraints, these are said to be
first class. If the constraints are first class, quantization can
proceed by requiring that the quantum operator version of the
constraints annihilate the wavefunctions.  Upon quantization, the
Poisson bracket algebra of constraints should translate into an
algebra of commutators where the commutator of any constraints should
again be expressible as a combination of the constraints. The
combination may involve coefficients that are functions of the
canonical variables (which become operators at a quantum level). It is
therefore crucial that such operators appear {\em to the left}. If
this were not the case, imposing that the wavefunctions be annihilated
by the constraints could imply ---via commutation relations--- extra
conditions on the wavefunctions.  The quantum theory found would
therefore not necessarily correspond to the classical theory one
started from.

{}From a physical point of view, constraints generate via canonical
transformations the symmetries of the theory. Therefore respecting the
constraint algebra at a quantum level is respecting the symmetries of
the theory under quantization. It should be noticed that imposing the
correct commutation relations between constraints at a quantum level
without anomalies is far from a mere technicality. For instance in the
case of string theory it is what determines that the theory of a
bosonic string only is consistent in 26 dimensions.

The canonical formulation of general relativity has constraints.
The issue of the commutation relations of the constraints
therefore appears for any quantized version of the theory that
one may consider. In this paper we will concentrate on the
attempt to quantize General Relativity based on the Loop
Representation \cite{RoSm} constructed from Ashtekar's new
variables \cite{As}. The issue that we want to address  is the
consistency of the constraints that have been proposed in
\cite{Ga,BrPu,foot1}. Although solutions to the proposed
quantum version of the constraints have been found \cite{RoSm,BrGaPu},
this does not imply consistency of the constraint algebra.

We will perform the calculation directly in the Loop
Representation. This will serve two purposes: firstly it shows
that the expressions presented in \cite{Ga,BrPu} are well defined
in the sense that commutators of the constraints can be
computed. Secondly it provides an excellent example of the use of
several techniques to perform computations in loop space. We have
made this paper deliberately detailed and explicit, at the risk of
being somewhat long, with the purpose of presenting in a clear
fashion the techniques used. These techniques can be applied to
other problems concerning the loop representation.

The problem of the constraint algebra in quantum gravity has received
a fair share of attention, both in terms of traditional
geometrodynamic variables and in terms of Ashtekar new variables. In
particular it is quite clear that the issue cannot be completely
analyzed on formal grounds: a suitable regularization is needed to
avoid ambiguities that can arise in the constraints.  A thorough
discussion of these and other issues, as well as a historical
introduction to the subject can be found in the papers by Tsamis and
Woodard \cite{TsWo} and Friedman and Jack \cite{FrJa}.

In spite of what we have just said, in this paper we will only discuss
the issue of the formal unregulated commutation relations. The
motivation is twofold. First of all, not even this calculation has
ever been performed in terms of loops. Second, our main purpose is
more to gain insight on how to operate with the recently introduced
expressions for the constraints in the loop representation rather than
to settle the issue of their consistency. The constraints in the loop
representation have been used to find solutions to the equations
\cite{BrGaPu}, but operating twice with a constraint requires
considerably more care than originally expected. Several lessons about
how to operate with the diffeomorphism and Hamiltonian constraints in
terms of loops will be learnt during the discussions of the different
commutators. A sharper understanding of how these operators act is
gained through this analysis.

This paper is organized as follows. In section II we will briefly
recall the definition of the constraints of canonical general
relativity and the
constraints of canonical quantum gravity in
the Loop Representation In section
III we compute the commutator of two diffeomorphism constraints,
in section IV the commutator of a diffeomorphism with a
Hamiltonian and in section V the commutator of two Hamiltonians.
Several technical aspects of the computations are discussed in the
appendices.

\section{Constraints in the loop representation}

There exist well established canonical formulations of general
relativity. The more traditional one is based on the use of a three
dimensional metric $q_{ab}$ and its conjugate momentum density
$\tilde{\pi}^{ab}$. Recently, a new formulation due to Ashtekar has
exhibited interesting properties. It is based on the use of an $SU(2)$
connection $A_a^i$ as fundamental variable and its conjugate momentum
density is a set of densitized triads $\tilde{E}^a_i$. In both
formulations the variables are not free but are subject to
constraints. These constraints group naturally into a vector and a
scalar. In terms of Ashtekar's new variables,
\begin{eqnarray}
C(\vec{N}) &=& \int d^3 x N^a(x) \tilde{E}^b_i(x) F_{ab}^i(x)\\
H(\ut{M}) &=& \int d^3 x \ut{M}(x) \tilde{E}^a_i(x) \tilde{E}^b_j(x)
F_{ab}^i(x) \epsilon^{ijk}
\end{eqnarray}
where $F_{ab}^i$ is the curvature of the Ashtekar connection.
It is important to remark that the Gauss constraint is not mentioned
because in the Loop representation it is satisfied automatically for
all loop functionals. We refer the reader to \cite{BrPu} for notation
and conventions.

In the loop representation of quantum gravity \cite{RoSm},
wavefunctions $\Psi(\gamma)$ are functions of loops $\gamma$. The word
loop has in this context a very precise meaning: it is an equivalence
class of unparametrized curves. The equivalence relation is that two
curves are equivalent if they give rise to the same holonomy for all
connections.  This means that two loops are equivalent if they only
differ by ``tails'' that immediately retrace themselves.

Because the theory started with an $SU(2)$ connection (the Ashtekar
connection), wavefunctions inherit certain $SU(2)$ identities
(Mandelstam identities),
\begin{eqnarray}
\Psi(\gamma) &=& \Psi(\gamma^{-1})\\
\Psi(\gamma_{1} \circ \gamma_{2}) &=& \Psi(\gamma_{2} \circ
\gamma_{1})\\
\Psi(\gamma_{1}\circ\gamma_{2}\circ\gamma_{3})+\
\Psi(\gamma_{1}\circ\gamma_{2}\circ\gamma_{3}^{-1})&=&
\Psi(\gamma_{2}\circ\gamma_{1}\circ\gamma_{3})+
\Psi(\gamma_{2}\circ\gamma_{1}\circ\gamma_{3}^{-1}).
\end{eqnarray}

In this representation, the diffeomorphism and Hamiltonian constraints
can be written by making use of the loop derivative. The loop
derivative is the derivative that arises in loop space when two loops
that differ by an infinitesimal element of area are considered close.
The definition of the loop derivative of a
function of a loop is,
\begin{equation}
\Delta_{ab}(\pi_{0}^{x}) \Psi(\gamma) = \lim_{\delta \gamma
\rightarrow
\cdot} {\Psi(\pi_{0}^{x}\circ \delta \gamma \circ \pi_{x}^{0}\circ
\gamma) -\Psi(\gamma) \over \sigma^{ab}}
\end{equation}
where $\delta \gamma$ is a loop of infinitesimal area
$\sigma^{ab}$ basepointed at $x$ and connected to the loop $\gamma$ ,
through a path $\pi_{0}^{x}$. We will usually
consider for practical computations infinitesimal loops formed by a
parallelogram of infinitesimal sides $\vec{u}$ and $\vec{v}$.

Notice that the loop derivative is a {\em path dependent} object. Loop
derivatives act on any loop or path dependence. In particular if one
computes the commutator of two loop derivatives there will be a
nonvanishing action of the second loop derivative on the path
dependence of the first loop derivative, leading to a commutation
relation of the form,
\begin{equation}
[ \Delta_{ab}(\pi_{o}^{x}) , \Delta_{cd}(\chi_{o}^{y})] =
\Delta_{ab}(\pi_{o}^{x})[\Delta_{cd}(\chi_{o}^{y})] =
-\Delta_{cd}(\chi_{o}^{y})[\Delta_{ab}(\pi_{o}^{x})]
\label{commutation}
\end{equation}
where the brackets denote that the second loop derivative only acts on
the loop dependence of whatever is in the brackets.

Another relation of importance is to notice that the loop derivative
satisfies a Bianchi identity. In order to define it, we need to
introduce another derivative in loop space, the covariant derivative.
The covariant derivative $D_{a}$ acts on functions of paths by
appending an infinitesimal element at the end of the path along the
coordinate direction $\hat{a}$,
\begin{equation}
(1+\epsilon u^{a} D_{a}) \Psi(\pi_{o}^{x}) =
\Psi(\pi_{o}^{x+\epsilon u}).
\end{equation}

With this definition, the Bianchi
identity reads,
\begin{equation}
D_{a}\Delta_{bc}(\pi_{o}^{x})+D_{b} \Delta_{ca}(\pi_{o}^{x}) +
D_{c} \Delta_{ab}(\pi_{o}^{x})=0.
\label{bianchi}
\end{equation}

The constraints of quantum gravity in the loop representation can be
obtained using the loop transform. They have been discussed in refs
\cite{Ga,BrPu} so we only discuss them briefly here to fix notation.

The diffeomorphism constraint is given by,
\begin{equation}
C(\vec{N}) \Psi(\gamma)= \int d^{3}x N^{a}(x) \oint_{\gamma} dy^{b}
\delta^{3}(x-y) \Delta_{ab}(\gamma_{o}^{y}) \Psi(\gamma)
\end{equation}
where $N^{a}$ is a vector field on the three manifold along which the
diffeomorphism is taken. This operator has been known for quite some
time to generate deformations of the loop in the argument of the
wavefunction corresponding to the diffeomorphism performed. As an
intuitive picture of this expression one should remember that in the
Ashtekar formulation of canonical gravity the diffeomorphism
constraint is given by $\int d^{3}x N^{a}(x) Tr[ E^{a}(x) F_{ab}(x)]$.
The expression in the loop representation can be heuristically thought
of as a replacement $E^{a}\rightarrow \oint_{\gamma} dy^{a}$ and
$F_{ab}\rightarrow \Delta_{ab}$. In this paper we will only consider
diffeomorphisms that leave unchanged the basepoint of the loops we
are using, that is, $\vec{N}$ vanishes at the basepoint.

The Hamiltonian constraint is,
\begin{equation}
H(\ut{M}) \Psi(\gamma)= \lim_{\epsilon\rightarrow 0}
\frac {\epsilon}{2} \int d^{3}x \ut{M}(x)
\oint_{\gamma} dy^{a} \oint_{\gamma} dz^{b} \delta^{3}(x-y)
f_{\epsilon}(y,y') [O_{y\,y'}\Delta_{ab}(\gamma_{o}^{y})
+O_{y\,o\,y'}\Delta_{ab}(\gamma_{o}^{y})]  \Psi(\gamma).
\end{equation}

In this paper we will omit the factor $\frac {1}{2}$.
This expression requires some discussion. First of all notice that we
have introduced a regulator $f_{\epsilon}(y,y')$ such that
$\lim_{\epsilon\rightarrow 0} f_{\epsilon}(y,y') \rightarrow
\delta^{3}(y,y')$. The need for this regulator can be seen directly
{}from the translation of the expression of the Hamiltonian constraint
in the Ashtekar formulation $\int d^{3}x M(x) \int d^{3}y
f_{\epsilon}(x,y) Tr[E^{a}(y) E^{b}(x) F_{ab}(x)]$ due to the fact
that the constraint is quadratic in momenta.  The expression is
naively zero if the loop is smooth (no kinks or intersections) due to
the fact that $\Delta_{ab}$ is antisymmetric.  One can check that when
the regulator is taken into account the expression also vanishes
provided it acts on diffeomorphism invariant functions.  Since
wavefunctions in the loop representation take values on all possible
loops we will generically consider a loop with a multiple intersection
at a given point (the case with more than one intersection is
trivially included since the action of the Hamiltonian is local).

A second point to notice is the appearance of the rerouting operator,
$O_{y\,y'}$. Acting on a function of a loop, this operator has the
following effect: it takes the loop formed by taking the original loop
between $y$ and $y'$ no matter the original orientation and then it
adds the portion between $y$ and $y'$ passing through the origin $o$,
no matter the original orientation. Of course, the points $y$ and $y'$
must lie on the loop, as they do in the expression of the Hamiltonian.
The reader may be puzzled at this point: if one cuts and pastes a
portion of a loop in the opposite orientation, one does not have in
general a loop any more. This is true. To complete the procedure for
arbitrary $y$ and $y'$ one must re-close the resulting loops (see
appendix A). However, in the limit where regulators are removed, the
points $y$ and $y'$ coincide with the intersection points and
therefore the action of the rerouting simply amounts to a rerouting of
certain ``petals'' of the loop. The ``re-tracings'' introduced to close
the loop are of higher order in powers of the regulator in the limit
where the regulators are removed.

The reader can realize that although applying these expressions to a
function of a loop seems reasonably straightforward, applying them for
a second time can introduce complications. For instance: what is the
action of two successive reroutings? What is the action of the loop
derivative on the rerouting? These are the kinds of questions we would
like to address in this paper.

\section{Diffeomorphism algebra}

We now proceed to compute the commutator of two diffeomorphism
constraints. The reader may feel this an unneeded computation.  After
all, if the constraint is the infinitesimal generator of
diffeomorphisms, it {\em should}, by definition, satisfy the correct
algebra. Although this is a valid viewpoint, we feel that an explicit
calculation is in order. Furthermore, the calculation can be viewed as
a confirmation that the proposed constraint is equivalent to the
generator of diffeomorphisms in loop space.  The calculation is quite
instructive in the sense that it makes a nontrivial use of several
properties of loop space.  This calculation is already present in the
literature \cite{GaTr} but we feel a more careful and detailed
derivation is in order.  We find it convenient to explain the
techniques used in some detail with the example of this calculation in
order to be able to be more succinct in the following ones.

We will start with an auxiliary calculation that will prove useful in
the sequel. We will evaluate the action of the loop derivative on a
diffeomorphism constraint. To this aim we construct the following
expression, which holds due to the very definition of loop derivative,

\begin{eqnarray}
(1+\sigma^{cd} \Delta_{cd}(\gamma_{o}^{z})) \int_{\gamma} dy^{a}
\delta(y-x) \Delta_{ab}(\gamma_{o}^{y}) \Psi(\gamma) := \nonumber\\
\int_{ \delta \gamma_{z} \circ \gamma}
dy^{a} \delta(y-x) \Delta_{ab}((\delta \gamma_{z} \circ
\gamma)_{o}^{y}) \Psi(\delta \gamma_{z} \circ \gamma).
 \label{defordiffeo}
\end{eqnarray}

In this expression, $\delta \gamma_{z}$ is the infinitesimal loop
added to $\gamma$ and connected trough a path from the origin up to
the point $z$.  That is, we evaluate the action of an infinitesimal
deformation of area $\sigma^{cd}$ acting on the infinitesimal
generator of diffeomorphisms. We now evaluate the right member of this
expression and this will enable us to read off the action of the loop
derivative on the infinitesimal generator of diffeomorphisms.

With this aim we expand the right member of (\ref{defordiffeo}),
partitioning the domain of integration, and using the definition of
the loop derivative to expand $\Psi(\delta \gamma\circ \gamma)$.
We get,
\begin{eqnarray}
&&\int_{o}^{z} dy^{a} \Delta_{ab}(\gamma_{o}^{y}) \delta(y-x)
(1+\sigma^{cd} \Delta_{cd}(\gamma_{o}^{z})) \Psi(\gamma)+\nonumber\\
&&+ \{ u^{a} \delta(z-x) \Delta_{ab}(\gamma_{o}^{z}) +v^{a}
\delta(z+u-x) \Delta_{ab}(\gamma_{o}^{z+u})
-u^{a} \delta(z+u+v-x) \Delta_{ab}(\gamma_{o}^{z+u+v})\nonumber\\
&&-v^{a} \delta(z+u-x) \Delta_{ab}(\gamma_{o}^{z+u+v+\bar{u}})\}
\quad (1+\sigma^{cd} \Delta_{cd}(\gamma_{o}^{z})) \Psi(\gamma)+
\nonumber\\
&&+\int_{z}^{1} dy^{a} \delta(y-x)
\Delta_{ab}((\delta \gamma_{z}\circ \gamma)_{o}^{y})
(1+\sigma^{cd} \Delta_{cd}(\gamma_{o}^{z}) \Psi(\gamma)\label{choclo}
\end{eqnarray}
where the terms in curly braces correspond to the integration along
the infinitesimal loop $\delta \gamma$ which we take as being given by
a parallelogram of infinitesimal sides given by $u^a$ and $v^b$.

The last term in this expression can be rewritten as
\begin{equation}
\oint_{\gamma} dy^{a} \delta(z-y) \Theta(z,y) (1+\sigma^{cd}
\Delta_{cd}(\gamma_{o}^{z}))[\Delta_{ab}(\gamma_{o}^{y})]
(1+\sigma^{ef}\Delta_{ef}(\gamma_{o}^{z})) \Psi(\gamma)
\end{equation}
where $\Theta(z,y)$ is a Heaviside function that orders the points
along the loop. We will be able to
combine the zeroth order contribution (in terms of the infinitesimal
loop) of this term with the first term in (\ref{choclo}). It should be
noticed that the first loop derivative does not act on everything to
its right but only on the path inside the second loop derivative
$\gamma_{0}^{y}$, a fact we denoted by enclosing it in brackets, and
that was discussed in detail in the previous section. We now consider
the expansion of the terms containing infinitesimally shifted loop.
They can be expressed with the use of the Mandelstam derivative,

\begin{eqnarray}
\Delta_{ab}(\gamma_{o}^{z+u}) &&= (1+u^{c} D_{c})
\Delta_{ab}(\gamma_{o}^{z})\\
\Delta_{ab}(\gamma_{o}^{z+u+v}) &&= (1+v^{d} D_{d}) (1+u^{c} D_{c})
\Delta_{ab}(\gamma_{o}^{z})\\
\delta(z+u-x) &&= (1+u^{a} \partial_{a}) \delta(z-x)
\end{eqnarray}

We now expand again (\ref{choclo})
\begin{eqnarray}
&&\oint_{\gamma} dy^{a} \delta(y-x) \Delta_{ab}(\gamma_{o}^{y})
(1+\sigma^{cd}\Delta_{cd}(\gamma_{o}^{z})) \Psi(\gamma)+\nonumber\\
&&+\{u^{a} \delta(z-x) \Delta_{ab}(\gamma_{o}^{z})+v^{a}
(1+u^{e}\partial_{e}) \delta(z-x) (1+u^{d}D_{d})
\Delta_{ab}(\gamma_{o}^{z}) -\nonumber\\
&&-u^{a} (1+v^{c}D_{c}) \delta(z-x) (1+v^{d} D_{d})
\Delta_{ab}(\gamma_{o}^{z}) -\nonumber\\
&&-v^{a}\delta(z-x) \Delta_{ab}(\gamma_{o}^{z})\}
(1+\sigma^{cd}\Delta_{cd}(\gamma_{o}^{z})) \Psi(\gamma)+\nonumber\\
&&+ \oint_{\gamma} dy^{a} \delta(y-x) \Theta(z,y) \sigma^{cd}
\Delta_{cd}(\gamma_{o}^{z})[\Delta_{ab}(\gamma_{o}^{y})]
(1+\sigma^{ef}\Delta_{ef}(\gamma_{o}^{z}))\Psi(\gamma).
\end{eqnarray}

Of the terms in braces, it can be readily seen that only
contributions proportional to $u^{a} v^{b}$ are present,
neglecting terms of higher order. The other terms combine to give
the original expression. We can finally read off the contribution
of the loop derivative,

\begin{eqnarray}\label{areadondiff}
&&\Delta_{cd}(\gamma_{o}^{z}) \oint_{\gamma}dy^{a} \delta(x-y)
\Delta_{ab}(\gamma_{o}^{y}) \Psi(\gamma) = \nonumber\\
&&2 [\partial_{[c} \delta(z-x)
\Delta_{d]b}(\gamma_{o}^{z})+\delta(z-x) D_{[c}
\Delta_{d]b}(\gamma_{o}^{z})] \Psi(\gamma)+\nonumber\\
&&\oint_{\gamma} dy^{a} \Theta(z,y) \delta(y-x)
\Delta_{cd}(\gamma_{o}^{z})[\Delta_{ab}(\gamma_{o}^{y})] \Psi(\gamma)
+ \nonumber\\
&&\oint_{\gamma} dy^{a}  \delta(y-x) \Delta_{ab}(\gamma_{o}^{y})
\Delta_{cd}(\gamma_{o}^{z}) \Psi(\gamma)\label{areadond}
\end{eqnarray}

With this calculation in hand, it is straightforward to compute
the successive action of two diffeomorphisms,

\begin{eqnarray}
&&C(\vec{N}) \, C(\vec{M}) \, \Psi(\gamma) = \\
&&\int d^{3}w N^{d}(w) \oint_{\gamma}
dz^{c} \delta(w-z) \Delta_{cd}(\gamma_{o}^{z}) \int d^{3}x M^{b}(x)
\oint_{\gamma} dy^{a} \delta(y-x) \Delta_{ab}(\gamma_{o}^{y})
\Psi(\gamma).
\nonumber
\end{eqnarray}
Expanding this expression, we get six terms,
\begin{eqnarray}
&&\int d^{3}w \int d^{3}x N^{d}(w) M^{b}(x) \oint_{\gamma} dz^{c}
\delta(w-z) \partial_{c} \delta(z-x) \Delta_{db}(\gamma_{o}^{z})
\Psi(\gamma)-\nonumber\\
&&\int d^{3}w \int d^{3}x N^{d}(w) M^{b}(x) \oint_{\gamma} dz^{c}
\delta(w-z) \partial_{d} \delta(z-x) \Delta_{cb}(\gamma_{o}^{z})
\Psi(\gamma)+\nonumber\\
&&\int d^{3}w \int d^{3}x N^{d}(w) M^{b}(x) \oint_{\gamma} dz^{c}
\delta(w-z) \delta(z-x) D_{c} \Delta_{d b}(\gamma_{o}^{z})
\Psi(\gamma)- \nonumber\\
&&\int d^{3}w \int d^{3}x N^{d}(w) M^{b}(x) \oint_{\gamma} dz^{c}
\delta(w-z) \delta(z-x) D_{d} \Delta_{c b}(\gamma_{o}^{z})
\Psi(\gamma)
+\nonumber\\
&&\int d^{3}w \int d^{3}x N^{d}(w) M^{b}(x) \oint_{\gamma} dz^{c}
\oint_{\gamma} dy^{a} \delta(w-z) \Theta(z,y) \delta(y-x)
\Delta_{cd}(\gamma_{o}^{z}) [\Delta_{ab}(\gamma_{o}^{y})] \Psi(\gamma)
+\nonumber\\
&&\int d^{3}w \int d^{3}x N^{d}(w) M^{b}(x) \oint_{\gamma} dz^{c}
\oint_{\gamma} dy^{a} \delta(w-z)  \delta(y-x)
\Delta_{ab}(\gamma_{o}^{y})\Delta_{cd}(\gamma_{o}^{z}) \Psi(\gamma)
\end{eqnarray}

We should now subtract the same terms with the replacement
$\vec{N} \leftrightarrow \vec{M}$. Since the calculation is
tedious but straightforward we describe in words how the terms
combine.  The fifth and sixth terms, when combined with the
similar terms coming from the substitution $\vec{N}
\leftrightarrow \vec{M}$ cancel taking into account the
commutation relations for the loop derivatives
(\ref{commutation}). The first and third term, combined with the
first of the substitution $\vec{N} \leftrightarrow \vec{M}$ form
a total derivative. The fourth term, combined with the third and
fourth of the substitution $\vec{N} \leftrightarrow \vec{M}$
cancel due to the Bianchi identities of the loop derivatives.
Finally, the second terms combine to produce exactly $C({\cal
L}_{\vec{N}} \vec{M})$, which is the correct result of the
calculation.

\section{Commutator of a diffeomorphism with a Hamiltonian}

This calculation will teach us about two important new ingredients
that were not present in the previous calculation: the action of
reroutings in the Hamiltonian and how to deal with
regularization. There has always been concerns about the use of a
background dependent regularization in the Hamiltonian constraint,
since it had potential to interfere with the action of
diffeomorphisms. With the help of this calculation it is possible to
detect the regularization problems explicitly, by showing where it is
needed to remove the regulators for the computation to work. The
commutator is,
\begin{eqnarray}
&&[C(\vec{M}) H(\ut{N})- H(\ut{N}) C(\vec{M})]=
\int d^{3}w M^{b}(w) \int d^{3}x \ut{N}(x) \times \nonumber\\
&&\oint_{L} dz^{a} \delta(z-w) \Delta_{ab}(\gamma_{o}^{z})
\oint_{\gamma} dy^{[c} \oint_{\gamma} dy'^{d]}\delta(x-y')
f_{\epsilon}(y-y')
[O_{y\,y'}+O_{y\,o\,y'}] \Delta_{cd}(\gamma_{o}^{y}) \Psi(\gamma)-
\nonumber\\
&&\oint_{\gamma} dy^{[c} \oint_{\gamma} dy'^{d]}\delta(x-y')
f_{\epsilon}(y-y')  [O_{y\,y'}+O_{y\,o\,y'}]
\Delta_{cd}(\gamma_{o}^{y})
\oint_{\gamma} dz^{a} \delta(z-w) \Delta_{ab}(\gamma_{o}^{z})
\Psi(\gamma).
\end{eqnarray}

We will now proceed as in the last section; first, we evaluate the
action of the loop derivative of the diffeomorphism constraint on the
loop derivative and the rerouting operators corresponding to the
Hamiltonian constraint. We will prove that these terms minus the
analogous ones coming from the action of the Hamiltonian on the
diffeomorphism cancel each other. The remaining terms, which are the
consequence of the action of the loop derivative on the loop
dependence of integrals will give rise to the correct
commutator. Since the terms involving $O_{y\,o\,y'}$ behave in an
analogous fashion to the ones that depend on $O_{y\,y'}$, we will only
concentrate on these.

The first contribution of the action of the diffeomorphism
on the Hamiltonian is,
\begin{eqnarray}
&&\oint_{\gamma} dz^{a} \delta(z-w)\oint_{\gamma} dy^{[c}
\oint_{\gamma} dy'^{d]}\delta(x-y')
f_{\epsilon}(y-y') \nonumber\\
 \Delta_{ab}(\gamma_{o}^{z})
[O_{y\,y'}\Delta_{cd}(\gamma_{o}^{y}) \Psi(\gamma)] .\nonumber
\end{eqnarray}

We now separate the outer integral into two portions, using the same
notation as in the rerouting operators, ie.  $\gamma_{y o y'}$ would
be the portion of the loop $\gamma$ that goes from $y$ to $y'$ through
the origin. We also perform a split of one of the inner integrals,
\begin{eqnarray}
&&\oint_{\gamma} dy^{[c}
\oint_{\gamma_{o\,y}} dy'^{d]} \biggr[ \oint_{\gamma_{y'\,y}} dz^{a}
+\oint_{\gamma_{y\,o\,y'}} dz^{a}\biggr]\delta(z-w)
\delta(x-y') f_{\epsilon}(y-y') \times \nonumber\\
&&\hspace{9.5cm} \times \Delta_{ab}(\gamma_{o}^{z})
[O_{y\,y'}\Delta_{cd}(\gamma_{o}^{y}) \Psi(\gamma)]+ \nonumber\\
&&+\oint_{\gamma} dy^{[c} \oint_{\gamma_{y\,o}} dy'^{d]}
\biggr[ \oint_{\gamma_{y\,y'}} dz^{a}
+\oint_{\gamma_{y'\,o\,y}} dz^{a}\biggr]\delta(z-w)\delta(x-y')
f_{\epsilon}(y-y') \times\nonumber\\
&&\hspace{9.5cm} \times \Delta_{ab}(\gamma_{o}^{z})
[O_{y\,y'}\Delta_{cd}(\gamma_{o}^{y}) \Psi(\gamma)].
\end{eqnarray}

After commuting the rerouting and the derivative (see appendix A),
and taking into account the sign introduced in this process we get,
\begin{eqnarray}
&&\oint_{\gamma} dy^{[c} \oint_{\gamma_{o\,y}} dy'^{d]}
\biggr[-\oint_{\gamma_{y'\,y}} dz^{a} +\oint_{\gamma_{y\,o\,y'}} dz^{a}
\biggr] \delta(z-w)\delta(x-y') f_{\epsilon}(y-y') \times
\nonumber\\
&&\hspace{9.5cm} \times O_{y\,y'}
\Delta_{ab}(\gamma_{o}^{z}) [\Delta_{cd}(\gamma_{o}^{y})
\Psi(\gamma)]+\nonumber\\
&&+\oint_{\gamma} dy^{[c} \oint_{\gamma_{y\,o}} dy'^{d]}
\biggr[\oint_{\gamma_{y\,y'}} dz^{a} -\oint_{\gamma_{y'\,o\,y}}
dz^{a}\biggr]
\delta(z-w)\delta(x-y') f_{\epsilon}(y-y') \times\nonumber\\
&&\hspace{9.5cm} \times O_{y\,y'} \Delta_{ab}(\gamma_{o}^{z})
[\Delta_{cd}(\gamma_{o}^{y}) \Psi(\gamma)].
\end{eqnarray}

The following step consists in noticing that the minus sign in front
of the integral can be absorbed by integrating along the loop in the
opposite direction, which we denote with an over-bar,
\begin{eqnarray}
&&\oint_{\gamma} dy^{[c} \oint_{\gamma_{o\,y}} dy'^{d]}
\biggr[\oint_{\overline{\gamma}_{y\,y'}} dz^{a}
+\oint_{\gamma_{y\,o\,y'}} dz^{a}\biggr]
\delta(z-w)\delta(x-y') f_{\epsilon}(y-y') \times\nonumber\\
&&\hspace{9.5cm} \times O_{y\,y'} \Delta_{ab}(\gamma_{o}^{z})
[\Delta_{cd}(\gamma_{o}^{y}) \Psi(\gamma)]+\nonumber\\
&&+\oint_{\gamma} dy^{[c} \oint_{\gamma_{y\,o}} dy'^{d]}
\biggr[\oint_{\gamma_{y\,y'}} dz^{a}
+\oint_{\overline{\gamma}_{y\,o\,y'}} dz^{a}\biggr]
\delta(z-w)\delta(x-y') f_{\epsilon}(y-y') \times\nonumber\\
&&\hspace{9.5cm} \times O_{y\,y'} \Delta_{ab}(\gamma_{o}^{z})
[\Delta_{cd}(\gamma_{o}^{y}) \Psi(\gamma)],
\end{eqnarray}
which in turns allows us to recombine the terms again into a single
loop integral, making use of the definition of the rerouting operator,
\begin{eqnarray}
&&
\oint_{\gamma} dy^{[c} \oint_{\gamma_{o\,y}} dy'^{d]} O_{y\,y'}
\oint_{\gamma} dz^{a}
\delta(z-w)\delta(x-y') f_{\epsilon}(y-y') \times
\Delta_{ab}(\gamma_{o}^{z})[\Delta_{cd}(\gamma_{o}^{y}) \Psi(\gamma)]
+\nonumber\\
&&+\oint_{\gamma} dy^{[c} \oint_{\gamma_{y\,o}} dy'^{d]} O_{y\,y'}
\oint_{\gamma} dz^{a}
\delta(z-w)\delta(x-y') f_{\epsilon}(y-y') \times
\Delta_{ab}(\gamma_{o}^{z}) [\Delta_{cd}(\gamma_{o}^{y}) \Psi(\gamma)].
\end{eqnarray}

So the final result of this manipulation can be written as,
\begin{eqnarray}
&&\oint_{\gamma} dy^{[c} \oint_{\gamma} dy'^{d]} O_{y\,y'}
\oint_{\gamma} dz^{a}
\delta(z-w)\delta(x-y') f_{\epsilon}(y-y') \times
\Delta_{ab}(\gamma_{o}^{z}) [\Delta_{cd}(\gamma_{o}^{y})
\Psi(\gamma)]=\nonumber\\
&&=\oint_{\gamma} dy^{[c} \oint_{\gamma} dy'^{d]}\delta(x-y')
f_{\epsilon}(y-y') \times O_{y\,y'}\nonumber\\
&&\oint_{\gamma} dz^{a}
\delta(z-w) [\Delta_{cd}(\gamma_{o}^{y}) \Delta_{ab}(\gamma_{o}^{z})
+ \Theta(z,y)\Delta_{ab}(\gamma_{o}^{z})[\Delta_{cd}(\gamma_{o}^{y})]]
\Psi(\gamma).\label{25}
\end{eqnarray}

The corresponding terms arising from the action of the Hamiltonian
constraint on the diffeomorphism are, \begin{eqnarray}
&&\oint_{\gamma} dy^{[c} \oint_{\gamma} dy'^{d]} O_{y\,y'}
\oint_{\gamma} dz^{a} \delta(z-w)\delta(x-y') f_{\epsilon}(y-y')
\times \Delta_{cd}(\gamma_{o}^{y}) [\Delta_{ab}(\gamma_{o}^{z})
\Psi(\gamma)]=\nonumber\\ &&=\oint_{\gamma} dy^{[c} \oint_{\gamma}
dy'^{d]} O_{y\,y'} \oint_{\gamma} dz^{a} \delta(z-w)\delta(x-y')
f_{\epsilon}(y-y') \times\nonumber\\ &&\times
[\Delta_{ab}(\gamma_{o}^{z}) \Delta_{cd}(\gamma_{o}^{y})+ \Theta(y,z)
\Delta_{cd}(\gamma_{o}^{y})[\Delta_{ab}(\gamma_{o}^{z})]]
\Psi(\gamma).\label{26} \end{eqnarray}

When we  subtract between (\ref{26}) from  (\ref{25}) one
immediately notices that the net result is zero if one uses the
commutation relation for the loop derivatives (\ref{commutation}).

We now concentrate on the terms that will give rise to the result
of the commutator. These terms appear when the
loop derivative acts on the loop dependence of the integrals,
exactly as in the commutator we considered in the previous section.

The contribution from $ C(\vec{M}) H(\ut{N}) \Psi(\gamma) $ is
\begin{eqnarray}
&&\int d^{3}w M^{b}(w) \oint_{\gamma} dz^{a}\delta(z-w)
\int d^{3}x \ut{N}(x) \times\nonumber\\
&&\hspace{2cm} \times \biggr(2 \oint_{\gamma}dy'^{d} \delta(x-y')
D_{[a}^{z} [f_{\epsilon}(z-y')  O_{z\,y'}\Delta_{b]d}(\gamma_{o}^{z})]
\Psi(\gamma)+\nonumber\\
&&\hspace{2.7cm} +2 \oint_{\gamma} dy^{c} D_{[b}^{z} [\delta(x-z)
f_{\epsilon}(y-z)
O_{y\,z}\Delta_{a]c}(\gamma_{o}^{y})] \Psi(\gamma) \biggr).
\end{eqnarray}

And the contribution from $ H(\ut{N}) C(\vec{M}) \Psi(\gamma) $ is
\begin{eqnarray}
&&\int d^{3}x \ut{N}(x) \int d^{3}w
M^{b}(w) \oint_{\gamma} dy^{[c} \oint_{\gamma}dy'^{d]} \delta(x-y')
f_{\epsilon}(y-y') \times\nonumber\\
&&\hspace{6cm} \times \biggr( 2 O_{y\,y'} D_{[c}^{y}
[\delta(y-w)\Delta_{d]b}(\gamma_{o}^{y})] \Psi(\gamma) \biggr).
\end{eqnarray}

The result of computing the commutator is,
\begin{eqnarray}
&&2 \oint_{\gamma} dz^{a}\delta(z-w)
\oint_{\gamma}dy'^{d} \delta(x-y')
D_{[a}^{z} [ f_{\epsilon}(z-y')  O_{z\,y'}
\Delta_{b]d}(\gamma_{o}^{z}) ]
\Psi(\gamma) +\\
&&+2 \oint_{\gamma} dz^{a}\delta(z-w)
\oint_{\gamma} dy^{c} D_{[b}^{z} (\delta(x-z) f_{\epsilon}(y-z)
O_{y\,z})\Delta_{a]c}(\gamma_{o}^{y})  \Psi(\gamma)
\nonumber\\
&&-\oint_{\gamma} dy^{[c} \oint_{\gamma}dy'^{d]} \delta(x-y')
f_{\epsilon}(y-y')
2 O_{y\,y'} D_{[c}^{y} [\delta(y-w)\Delta_{d]b}(\gamma_{o}^{y})]
\Psi(\gamma). \nonumber
\end{eqnarray}

We now explicitly write the antisymmetrizations of the first two
terms,and use Leibnitz' rule in the last one we get,
\begin{eqnarray}\label{fiveterms}
&& \int d^{3}w M^{b}(w) \int d^{3}x \ut{N}(x) \times \nonumber\\
\times && \biggr( \oint_{\gamma} dz^{a}\delta(z-w)
 \oint_{\gamma}dy'^{d} \delta(x-y')
D_{a}^{z} [ f_{\epsilon}(z-y')  O_{z\,y'}\Delta_{bd}(\gamma_{0}^{z})]
\Psi(\gamma) -\nonumber\\
-&& \oint_{\gamma} dz^{a}\delta(z-w)  \oint_{\gamma}dy'^{d}
\delta(x-y')
D_{b}^{z} [ f_{\epsilon}(z-y') O_{z\,y'} \Delta_{ad}(\gamma_{0}^{z})]
\Psi(\gamma) +\nonumber\\
+&& \oint_{\gamma} dz^{a}\delta(z-w)
\oint_{\gamma} dy^{c} D_{b}^{z} (\delta(x-z) f_{\epsilon}(y-z)
O_{y\,z})\Delta_{ac}(\gamma_{0}^{y})  \Psi(\gamma) -\\
-&&  \oint_{\gamma} dz^{a}\delta(z-w)
\oint_{\gamma} dy^{c} D_{a}^{z} (\delta(x-z) f_{\epsilon}(y-z)
O_{y\,z})\Delta_{bc}(\gamma_{0}^{y})  \Psi(\gamma)
\nonumber\\
&&\noindent -\oint_{\gamma} dy^{[c} \oint_{\gamma}dy'^{d]}
\delta(x-y')
f_{\epsilon}(y-y')
 O_{y\,y'} [\partial_{c}^{y}\delta(y-w)\Delta_{db}(\gamma_{o}^{y})
-\partial_{d}^{y}\delta(y-w)\Delta_{cb}(\gamma_{o}^{y})] \Psi(\gamma)
\nonumber\\
&&\noindent -\oint_{\gamma} dy^{[c} \oint_{\gamma}dy'^{d]} \delta(x-y')
f_{\epsilon}(y-y')
 O_{y\,y'} \delta(y-w)[D_{c}^{y}\Delta_{db}(\gamma_{o}^{y})
-D_{d}^{y}\Delta_{cb}(\gamma_{o}^{y})] \Psi(\gamma)\biggr).
\nonumber
\end{eqnarray}

The first and fourth terms can be integrated by parts on $z$.  Taking
into account that the diffeomorphisms we are considering have trivial
action at the basepoint of the loops, the resulting terms cancel with
the fifth term. In order for this cancelation to occur we have to
remove the regulator so that
$\partial_{d}^{y'}\delta(y'-w)f_{\epsilon}(y-y')=
\partial_{d}^{y}\delta(y-w)f_{\epsilon}(y-y')$. The remaining
terms are the second and third. By using the Bianchi identity in the
last term of (\ref{fiveterms}), this equation may be rewritten as
\begin{eqnarray}\label{20}
&& \int d^{3}w d^{3}x M^{b}(w)  \ut{N}(x)\times\nonumber\\
\times&&\biggr(
- \oint_{\gamma} dy^{c}\delta(y-w)  \oint_{\gamma}dy'^{d} \delta(x-y')
D^{y}_{b} [ f_{\epsilon}(y-y') O_{y\,y'} \Delta_{cd}(\gamma_{o}^{y})]
\Psi(\gamma) +\nonumber\\
&&+ \oint_{\gamma} dy^{c}\delta(y-w)
\oint_{\gamma} dy'^{d} D^{y}_{b} (\delta(x-y) f_{\epsilon}(y'-y)
O_{y'\,y})\Delta_{cd}(\gamma_{o}^{y'})  \Psi(\gamma) +\nonumber\\
&&+\oint_{\gamma} dy^{[c} \oint_{\gamma}dy'^{d]} \delta(y-w)\delta(x-y')
f_{\epsilon}(y-y')
 O_{y\,y'} D_{b}^{y}\Delta_{cd}(\gamma_{o}^{y})  \Psi(\gamma)\biggr).
\end{eqnarray}

By using Leibnitz rule in the first term we get,
\begin{eqnarray}\label{21}
&& \int d^{3}w d^{3}x M^{b}(w)  \ut{N}(x)\times\nonumber\\
&&\times\biggr(
- \oint_{\gamma} dy^{c}\delta(y-w)  \oint_{\gamma}dy'^{d} \delta(x-y')
D^{y}_{b} [ f_{\epsilon}(y-y') O_{y\,y'}] \Delta_{cd}(\gamma_{o}^{y})
\Psi(\gamma) +\nonumber\\
&&+ \oint_{\gamma} dy^{c}\delta(y-w)
\oint_{\gamma} dy'^{d} D^{y}_{b} (\delta(x-y) f_{\epsilon}(y'-y)
O_{y'\,y})\Delta_{cd}(\gamma_{o}^{y'})  \Psi(\gamma)\biggr).\label{a1}
\end{eqnarray}

Taking into account the definition of  the  rerouting
operator, we have,
\begin{equation}\label{22}
D^{y}_{b}[f_{\epsilon}(y-y')O_{y\,y'}] =
D^{y}_{b}f_{\epsilon}(y-y')
O_{y\,y'}  = -D^{y'}_{b}f_{\epsilon}(y-y')O_{y\,y'}
\end{equation}
and using the Leibnitz rule again in (\ref{a1}) we get
\begin{eqnarray}\label{23}
&& \int d^{3}w d^{3}x M^{b}(w)  \ut{N}(x)\times\nonumber\\
&&\times\biggr(+ \oint_{\gamma} dy^{c}\oint_{\gamma}dy'^{d}
D^{y'}_{b}[\delta(y-w)  \delta(x-y')
 f_{\epsilon}(y-y') O_{y\,y'} \Delta_{cd}(\gamma_{o}^{y})]
\Psi(\gamma) +\nonumber\\
&&+ \oint_{\gamma} dy^{c}
\oint_{\gamma} dy'^{d} D^{y}_{b} [\delta(y-w)\delta(x-y)
f_{\epsilon}(y'-y)
O_{y'\,y}\Delta_{cd}(\gamma_{o}^{y'})]  \Psi(\gamma) +\nonumber\\
&&- \oint_{\gamma} dy^{c}\delta(y-w)  \oint_{\gamma}dy'^{d}
\partial^{y'}_{b} \delta(x-y')
 f_{\epsilon}(y-y') O_{y\,y'} \Delta_{cd}(\gamma_{o}^{y})
\Psi(\gamma) +\nonumber\\
&&- \oint_{\gamma} dy^{c}\partial^{y}_{b}\delta(y-w)
\oint_{\gamma} dy'^{d}  \delta(x-y) f_{\epsilon}(y'-y)
O_{y'\,y}\Delta_{cd}(\gamma_{o}^{y'})  \Psi(\gamma).
\end{eqnarray}

The second term cancels with the first after interchanging  $y$
and $y'$ (and again removing the regulator so that $ \delta(y-w)
f_{\epsilon}(y-y')=\delta(y'-w) f_{\epsilon}(y-y')$). If we
interchange $y$ and $y'$ in the last term  and  integrate  by parts
in $x$ and $w$ we get the final result,
\begin{eqnarray}
&&\int d^{3}w d^{3}x  (\ut{N}(x) \partial_{b} M^{b}(w)
-M^{b}(w)  \partial_{b} \ut{N}(x))
\oint_{\gamma} dz^{a} \oint_{\gamma}dy'^{d}  \times\nonumber\\
&&\times \delta(z-w) \delta(x-y')
f_{\epsilon}(z-y')  \Delta_{ad}(\gamma_{o}^{z})  O_{z\,y'} \Psi(\gamma)
= H({\cal L}_{\vec M} \ut{N})
\end{eqnarray}
which reproduces the classical commutator.

\section{Commutator of two Hamiltonians}

The commutator of two Hamiltonians introduces another new
computational requirement. While computing the commutator, one gets
the product of two rerouting operators: one per each Hamiltonian.
The result one expects involves only one rerouting, the one due to the
presence of the metric in the function smearing the resulting
diffeomorphism. We will give greater details of this aspect of the
calculation in appendix B.

The commutator of two Hamiltonians can be written as,
\begin{eqnarray}
&&[H(\ut{N}) H(\ut{M})-H(\ut{M}) H(\ut{N})] \Psi(L)=
\int d^{3}x \ut{N}(x)\int d^{3}w \ut{M}(w)\times\nonumber\\
&&\oint_{\gamma} dy^{[c}\oint_{\gamma} dy'^{d]} \delta(x-y')
f_{\epsilon}(y-y') [O_{y\,y'}+O_{y\,o\,y'}] \Delta_{cd}(\gamma_{o}^{y})
\times\nonumber\\&& \oint_{\gamma} dz^{[p}\oint_{\gamma} dz'^{h]}
\delta(w-z') f_{\rho}(z-z') [O_{z\,z'}+O_{z\,o\,z'}]
\Delta_{ph}(\gamma_{o}^{z})\Psi(\gamma)-\nonumber\\
&&-\int d^{3}x \ut{N}(x) \int d^{3}(w) \ut{M}(w)\times\nonumber\\
&& \oint_{\gamma} dz^{[p}\oint_{\gamma} dz'^{h]}
\delta(w-z') f_{\rho}(z-z') [O_{z\,z'}+O_{z\,o\,z'}]
\Delta_{ph}(\gamma_{o}^{z})\times\nonumber\\
&&\oint_{\gamma} dy^{[c}\oint_{\gamma} dy'^{d]} \delta(x-y')
f_{\epsilon}(y-y')
[O_{y\,y'}+O_{y\,o\,y'}] \Delta_{cd}(\gamma_{o}^{y})\Psi(\gamma).
\label{42}
\end{eqnarray}

We now proceed as in the previous case. Since several terms  in the
commutator cancel in similar fashion, we will show the cancellation
explicitly for one case and will indicate how to proceed with the
others verbally.

Let us start by considering the terms that arise from the action of
the derivative operator of the first Hamiltonian on the integrand of
the second operator. We will discuss later on the terms that arise
{}from the action of the derivative operator on the loop dependence of
the integrals of the second Hamiltonian, which will give rise to the
nonvanishing part of the commutator, which reproduces the classical
result.

The calculation will be performed as follows. We will consider each of
the above mentioned terms in $H(\ut{N}) H(\ut{M})$ and perform
operations that show that they are actually term in $H(\ut{N})
H(\ut{M})$ and therefore cancel in the commutator. This will involve
proving identities among integrals along petals of intersecting loops.
For this it is convenient to introduce loop diagrams. In these
diagrams we represent each loop as an oriented segment (from left to
right) and we mark the ordering of the different variables in the
integrands along the loops. We will see that in several cases the
order in which variables appear implies that the integrals are
actually zero. Several operations we will perform involving the
rerouting operator will resemble the ones we performed in the previous
section.

We now consider one of the contributions indicated above, extract the
rerouting operator outside the loop integral as we did in the previous
section and we get,
\begin{eqnarray}
&&\oint_{\gamma} dy^{[c}\oint_{\gamma_{o\,y}} dy'^{d]} \delta(x-y')
f_{\epsilon}(y-y')
\times\nonumber\\
&& \oint_{[O_{y\,y'}\gamma]_{o\,y'}} dz^{[p}
\oint_{[O_{y\,y'}\gamma]_{y\,o}} dz'^{h]}
\delta(w-z') f_{\rho}(z-z')O_{y\,y'}\Delta_{cd}(\gamma_{o}^{y})
[O_{z\,z'}\Delta_{ph}(\gamma_{o}^{z})\Psi(\gamma)]+\nonumber\\
&&+\oint_{\gamma} dy^{[c}\oint_{\gamma_{y\,o}} dy'^{d]} \delta(x-y')
f_{\epsilon}(y-y') \times\nonumber\\
&& \oint_{[O_{y\,o\,y'}\gamma]_{o\,y'}} dz^{[p}
\oint_{[O_{y\,o\,y'}\gamma]_{y\,o}} dz'^{h]}
\delta(w-z') f_{\rho}(z-z')O_{y\,o\,y'}\Delta_{cd}(\gamma_{o}^{y})
[O_{z\,z'}\Delta_{ph}(\gamma_{o}^{z})\Psi(\gamma)]=\nonumber\\
&&=\oint_{\gamma} dy^{[c}\oint_{\gamma_{o\,y}} dy'^{d]} \delta(x-y')
f_{\epsilon}(y-y') \times\nonumber\\
&& \oint_{[O_{y\,y'}\gamma]_{o\,y'}} dz^{[p}
\oint_{[O_{y\,y'}\gamma]_{y\,o}} dz'^{h]}
\delta(w-z') f_{\rho}(z-z')O_{y\,y'}
O_{z\,z'} \Delta_{cd}(\gamma_{o}^{y}) [\Delta_{ph}(\gamma_{o}^{z})
\Psi(\gamma)]-\nonumber\\
&&-\oint_{\gamma} dy^{[c}\oint_{\gamma_{y\,o}} dy'^{d]} \delta(x-y')
f_{\epsilon}(y-y') \times\nonumber\\
&& \oint_{[O_{y\,o\,y'}\gamma]_{o\,y'}} dz^{[p}
\oint_{[O_{y\,o\,y'}\gamma]_{y\,o}} dz'^{h]}
\delta(w-z') f_{\rho}(z-z')O_{y\,o\,y'}
O_{z\,z'} \Delta_{cd}(\gamma_{o}^{y})[\Delta_{ph}(\gamma_{o}^{z})
\Psi(\gamma)]. \label{a3}
\end{eqnarray}

We now use the commutation relation for loop derivatives
(\ref{commutation}) to obtain the identity,
\begin{eqnarray}
&&\Delta_{cd}(\gamma_{o}^{y}) [\Delta_{ph}(\gamma_{o}^{z})
\Psi(\gamma)]
=\nonumber\\
&&=[\Delta_{p\,h}(\gamma_{o}^{z})\ \Delta_{c\,d}(\gamma_{o}^{y})+
\theta(z-y)
\Delta_{c\,d}(\gamma_{o}^{y})[\Delta_{p\,h}(\gamma_{o}^{z})]]
\Psi(\gamma)=
\nonumber\\
&&=[\Delta_{p\,h}(\gamma_{o}^{z})[\Delta_{c\,d}(\gamma_{o}^{y})]+
\Delta_{c\,d}(\gamma_{o}^{y})\Delta_{p\,h}(\gamma_{o}^{z})-
\Theta(y,z)\Delta_{p\,h}(\gamma_{o}^{z})
[\Delta_{c\,d}(\gamma_{o}^{y})]]
\Psi(\gamma)= \nonumber\\
&&=[\Theta(z,y)
\Delta_{p\,h}(\gamma_{o}^{z}) [\Delta_{c\,d}(\gamma_{o}^{y})]+
\Delta_{c\,d}(\gamma_{o}^{y})  \Delta_{p\,h}(\gamma_{o}^{z})]
\Psi(\gamma)=
\nonumber\\
&&=\Delta_{p\,h}(\gamma_{o}^{z})[\Delta_{c\,d}(\gamma_{o}^{y})
\Psi(\gamma)].
\end{eqnarray}

It is now useful to draw the loop diagrams corresponding to the
integrals present in (\ref{a3}). In the diagrams we indicate the
relative domain of integration of the different variables in the
first term of the expression,
\\ \\
\unitlength=1.00mm
\linethickness{0.4pt}
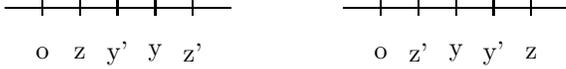
\begin{picture}(95.00,16.00)
\put(20.00,15.00){\line(1,0){5.00}}
\put(25.00,15.00){\line(0,1){1.00}}
\put(25.00,16.00){\line(0,-1){2.00}}
\put(25.00,14.00){\line(0,1){1.00}}
\put(25.00,15.00){\line(1,0){5.00}}
\put(30.00,15.00){\line(0,1){1.00}}
\put(30.00,16.00){\line(0,-1){2.00}}
\put(30.00,14.00){\line(0,1){1.00}}
\put(30.00,15.00){\line(1,0){5.00}}
\put(35.00,15.00){\line(0,1){1.00}}
\put(35.00,16.00){\line(0,-1){2.00}}
\put(35.00,14.00){\line(0,1){1.00}}
\put(35.00,15.00){\line(1,0){5.00}}
\put(40.00,15.00){\line(0,1){1.00}}
\put(40.00,16.00){\line(0,-1){2.00}}
\put(40.00,14.00){\line(0,1){1.00}}
\put(40.00,15.00){\line(1,0){5.00}}
\put(45.00,15.00){\line(0,1){1.00}}
\put(45.00,16.00){\line(0,-1){2.00}}
\put(45.00,14.00){\line(0,1){1.00}}
\put(45.00,15.00){\line(1,0){5.00}}
\put(65.00,15.00){\line(1,0){5.00}}
\put(70.00,15.00){\line(0,1){1.00}}
\put(70.00,16.00){\line(0,-1){2.00}}
\put(70.00,14.00){\line(0,1){1.00}}
\put(70.00,15.00){\line(1,0){5.00}}
\put(75.00,15.00){\line(0,1){1.00}}
\put(75.00,16.00){\line(0,-1){2.00}}
\put(75.00,14.00){\line(0,1){1.00}}
\put(75.00,15.00){\line(1,0){5.00}}
\put(80.00,15.00){\line(0,1){1.00}}
\put(80.00,16.00){\line(0,-1){2.00}}
\put(80.00,14.00){\line(0,1){1.00}}
\put(80.00,15.00){\line(1,0){5.00}}
\put(85.00,15.00){\line(0,1){1.00}}
\put(85.00,16.00){\line(0,-1){2.00}}
\put(85.00,14.00){\line(0,1){1.00}}
\put(85.00,15.00){\line(1,0){5.00}}
\put(90.00,15.00){\line(0,1){1.00}}
\put(90.00,16.00){\line(0,-1){2.00}}
\put(90.00,14.00){\line(0,1){1.00}}
\put(90.00,15.00){\line(1,0){5.00}}
\put(25.00,9.00){\makebox(0,0)[cc]{o}}
\put(30.00,9.00){\makebox(0,0)[cc]{z}}
\put(35.00,9.00){\makebox(0,0)[cc]{y'}}
\put(40.00,9.00){\makebox(0,0)[cc]{y}}
\put(70.00,9.00){\makebox(0,0)[cc]{o}}
\put(75.00,9.00){\makebox(0,0)[cc]{z'}}
\put(80.00,9.00){\makebox(0,0)[cc]{y}}
\put(85.00,9.00){\makebox(0,0)[cc]{y'}}
\put(90.00,9.00){\makebox(0,0)[cc]{z}}
\put(45.00,9.00){\makebox(0,0)[cc]{z'}}
\end{picture}

By looking at the diagram, we see we can rewrite the integrals in
the following way
\begin{eqnarray}
&&\oint_{\gamma} dz^{[p}\oint_{\gamma_{z\,o}} dz'^{h]}
\delta(w-z') f_{\rho}(z-z') \times\nonumber\\
&&\oint_{[O_{z\,z'}\gamma]_{z\,z'}} dy^{[c}
\oint_{[O_{z\,z'}\gamma]_{z\,y}}
dy'^{d]} \delta(x-y') f_{\epsilon}(y-y')
O_{y\,y'} O_{z\,z'} \Delta_{ph}(\gamma_{o}^{z})
[\Delta_{cd}(\gamma_{o}^{y})\Psi(\gamma)]-\nonumber\\
&&-\oint_{\gamma} dz^{[p}\oint_{\gamma_{o\,z}} dz'^{h]}
\delta(w-z') f_{\rho}(z-z') \times\nonumber\\
&&\oint_{[O_{z\,z'}\gamma]_{z\,z'}}
dy^{[c}\oint_{[O_{z\,z'}\gamma]_{z\,y}}
dy'^{d]} \delta(x-y') f_{\epsilon}(y-y')
O_{y\,o\,y'} O_{z\,z'}\Delta_{ph}(\gamma_{o}^{z})
[\Delta_{cd}(\gamma_{o}^{y})\Psi(\gamma)].
\end{eqnarray}

We now use the fact that $ O_{y\,y'} O_{z\,z'}=O_{z\,z'} O_{y\,y'}$
and commute rerouting and derivative operators again.
\begin{eqnarray}
&&\oint_{\gamma} dz^{[p}\oint_{\gamma_{z\,o}} dz'^{h]}
\delta(w-z') f_{\rho}(z-z') \times\nonumber\\
&&\oint_{[O_{z\,z'}\gamma]_{z\,z'}}
dy^{[c}\oint_{[O_{z\,z'}\gamma]_{z\,y}}
dy'^{d]} \delta(x-y') f_{\epsilon}(y-y')
O_{z\,z'} O_{y\,y'} \Delta_{ph}(\gamma_{o}^{z})
[\Delta_{cd}(\gamma_{o}^{y})\Psi(\gamma)]-\nonumber\\
&&-\oint_{\gamma} dz^{[p}\oint_{\gamma_{o\,z}} dz'^{h]}
\delta(w-z') f_{\rho}(z-z') \times\nonumber\\
&&\oint_{[O_{z\,z'}\gamma]_{z\,z'}}
dy^{[c}\oint_{[O_{z\,z'}\gamma]_{z\,y}}
dy'^{d]} \delta(x-y') f_{\epsilon}(y-y')
O_{z\,z'} O_{y\,o\,y'} \Delta_{ph}(\gamma_{o}^{z})
[\Delta_{cd}(\gamma_{o}^{y})\Psi(\gamma)]=\nonumber\\
&&=\oint_{\gamma} dz^{[p}\oint_{\gamma_{z\,o}} dz'^{h]}
\delta(w-z') f_{\rho}(z-z') \times\nonumber\\
&&\oint_{[O_{z\,z'}\gamma]_{z\,z'}}
dy^{[c}\oint_{[O_{z\,z'}\gamma]_{z\,y}}
dy'^{d]} \delta(x-y') f_{\epsilon}(y-y')
O_{z\,z'} \Delta_{ph}(\gamma_{o}^{z}) [O_{y\,y'}
\Delta_{cd}(\gamma_{o}^{y})\Psi(\gamma)]+\nonumber\\
&&+\oint_{\gamma} dz^{[p}\oint_{\gamma_{o\,z}} dz'^{h]}
\delta(w-z') f_{\rho}(z-z') \times\nonumber\\
&&\oint_{[O_{z\,z'}\gamma]_{z\,z'}}
dy^{[c}\oint_{[O_{z\,z'}\gamma]_{z\,y}}
dy'^{d]} \delta(x-y') f_{\epsilon}(y-y')
O_{z\,z'} \Delta_{ph}(\gamma_{o}^{z}) [O_{y\,o\,y'}
\Delta_{cd}(\gamma_{o}^{y})\Psi(\gamma)]=\nonumber\\
&&=\oint_{\gamma} dz^{[p}\oint_{\gamma} dz'^{h]}
\delta(w-z') f_{\rho}(z-z') \times\nonumber\\
&&\oint_{[O_{z\,z'}\gamma]_{z\,z'}}
dy^{[c}\oint_{[O_{z\,z'}\gamma]_{z\,y}}
dy'^{d]} \delta(x-y') f_{\epsilon}(y-y')
O_{z\,z'} \Delta_{ph}(\gamma_{o}^{z})[[O_{y\,y'}+O_{y\,o\,y'}]
\Delta_{cd}(\gamma_{o}^{y})\Psi(\gamma)].
\end{eqnarray}

This proves that all terms with four loop integrals that appear
in the commutator (\ref{42}) cancel among themselves
(each order of the operators in the commutator produces an equal
contribution).

We now need to address the terms with three loop integrals.  In order
to do this we begin with the following observation. If in the second
term of expression (\ref{42}) we rename pairwise the names of the
following variables, $z$ and $y$, $z'$ and $y'$, $p$ and $c$ and
finally $h$ and $d$. The only real change in the last term of the
commutator is that $x$ gets replaced by $w$ in the delta's arguments.
Then, we may work with the first term and only at the end add the
contribution of the second, which is similar in form. As before, we
need only consider the proof for $O_{y\,y'} O_{z\,z'}$. Since in the
terms we are considering the loop derivative acts only on the loop
dependence of the integrals, terms with different reroutings can be
treated in a similar fashion.

We begin studying the action of the loop derivative of the
first Hamiltonian on the third and fourth loop integrals,
\begin{eqnarray}
&&\oint_{\delta \gamma_{y}o\gamma} dz^{[p}\oint_{\gamma} dz'^{h]}
\delta(w-z')
f_{\rho}(z-z')
O_{z\,z'} \Delta_{ph}(\gamma_{o}^{z}) \Psi(\gamma) -\nonumber\\
&&-\oint_{\gamma} dz^{[p}\oint_{\gamma} dz'^{h]}
\delta(w-z') f_{\rho}(z-z')
O_{z\,z'} \Delta_{ph}(\gamma_{o}^{z}) \Psi(\gamma)+\nonumber\\
&&+\oint_{\gamma} dz^{[p}\oint_{\delta \gamma_{y}o\gamma} dz'^{h]}
\delta(w-z')
f_{\rho}(z-z')O_{z\,z'} \Delta_{ph}(\gamma_{o}^{z}) \Psi(\gamma)
\nonumber\\
&&-\oint_{\gamma} dz^{[p}\oint_{\gamma} dz'^{h]} \delta(w-z')
f_{\rho}(z-z')
O_{z\,z'} \Delta_{ph}(\gamma_{o}^{z}) \Psi(\gamma).
\end{eqnarray}

In the first integral we need to do the following substitution
(valid only in the limit when the regulator is removed),
\begin{eqnarray}
f_{\rho}(z-z')O_{z\,z'} \Delta_{ph}(\gamma_{o}^{z}) \Psi(\gamma)=
-f_{\rho}(z-z')O_{z'\,o\,z} \Delta_{ph}(\gamma_{o}^{z'}). \label{31}
\end{eqnarray}
As an aid  to understand this expression one should remember that the
corresponding expression in the connection representation would be,
\begin{eqnarray}
f_{\rho}(z-z') Tr[F_{ph}(z)U_{z\,z'}U_{z\,o\,z'}]
=- f_{\rho}(z-z') Tr[F_{ph}(z')U_{z'\,o\,z}U_{z'\,z}]
\end{eqnarray}

Finally, working in an analogous way as in the previous section the
result is,
\begin{eqnarray}
&&\oint_{\gamma} dy^{[c}\oint_{\gamma} dy'^{d]} \delta(x-y')
f_{\epsilon}(y-y')\times\nonumber\\
&&O_{y\,y'}\left( 2\oint_{\gamma} dz^{p}
\delta(w-z)D_{[c}^{y}[f_{\rho}(y-z)
O_{z\,o\,y}\Delta_{d]p}(\gamma_{o}^{z})]\Psi(\gamma) + \right.
\nonumber\\
&&\left.+2\oint_{\gamma} dz^{p}D_{[c}^{y}[\delta(w-y)f_{\rho}(y-z)
O_{z\,y}\Delta_{d]p}(\gamma_{o}^{z})] \Psi(\gamma)\right).
\end{eqnarray}

Considering the fact that the Mandelstam derivative
commutes with the loop integral
and the rerouting operator, as it can be seen  through the results of
appendix A, we get,
\begin{eqnarray}
&&\oint_{\gamma} dy^{c}\oint_{\gamma} dy'^{d} \delta(x-y')
f_{\epsilon}(y-y')\times\nonumber\\
&&\biggr(D_{c}^{y}[O_{y\,y'}
\oint_{\gamma} dz^{p}[\delta(w-z)f_{\rho}(y-z)
O_{z\,o\,y}+\delta(w-y)f_{\rho}(z-y)O_{z\,y}]
\Delta_{dp}(\gamma_{o}^{z})]\Psi(\gamma) -\nonumber\\
&&-D_{d}^{y}[O_{y\,y'} \oint_{\gamma} dz^{p}[\delta(w-z)f_{\rho}(y-z)
O_{z\,o\,y}+\delta(w-y)f_{\rho}(z-y)O_{z\,y}]
\Delta_{cp}(\gamma_{o}^{z})]\Psi(\gamma)\biggr),\nonumber\\
\end{eqnarray}
and removing the regulator we get,
\begin{eqnarray}
&&\oint_{\gamma} dy^{c}\oint_{\gamma} dy'^{d} \delta(x-y')
f_{\epsilon}(y-y')\times\nonumber\\
&&\biggr(\partial_{c}^{y}\delta(w-y)O_{y\,y'}
\oint_{\gamma} dz^{p}[f_{\rho}(y-z)
O_{z\,o\,y}+f_{\rho}(z-y)O_{z\,y}]
\Delta_{dp}(\gamma_{o}^{z})]\Psi(\gamma)+\nonumber\\
&&+\delta(w-y)D_{c}^{y}[O_{y\,y'} \oint_{\gamma} dz^{p}[f_{\rho}(y-z)
O_{z\,o\,y}+f_{\rho}(z-y)O_{z\,y}]\Delta_{dp}(\gamma_{o}^{z})]
\Psi(\gamma)
-\nonumber\\
&&-\partial_{d}^{y}\delta(w-y)O_{y\,y'}
\oint_{\gamma} dz^{p}[f_{\rho}(y-z)
O_{z\,o\,y}+f_{\rho}(z-y)O_{z\,y}]\Delta_{cp}(\gamma_{o}^{z})]
\Psi(\gamma)-\nonumber\\
&&-\delta(w-y)D_{d}^{y}[O_{y\,y'} \oint_{\gamma} dz^{p}[f_{\rho}(y-z)
O_{z\,o\,y}+f_{\rho}(z-y)O_{z\,y}]\Delta_{cp}(\gamma_{o}^{z})]
\Psi(\gamma)
\biggr)
\end{eqnarray}

In the limit in which the regulator is removed the product
$\delta(x-y)\times\delta(w-y) $
is symmetric in $x$ and $w$ and therefore some of the terms in the
above expression cancel with those coming from the other term in the
commutator. The remaining terms are
\begin{eqnarray}
&&\oint_{\gamma} dy^{c}\oint_{\gamma} dy'^{d} \delta(x-y')
f_{\epsilon}(y-y')\times\nonumber\\
&&(-\partial_{c}^{w}\delta(w-y))O_{y\,y'}
\oint_{\gamma} dz^{p}[f_{\rho}(y-z)
O_{z\,o\,y}+f_{\rho}(z-y)O_{z\,y}]\Delta_{dp}(\gamma_{o}^{z})]
\Psi(\gamma)-\nonumber\\
&&-\oint_{\gamma} dy^{c}\oint_{\gamma} dy'^{d} \delta(x-y')
f_{\epsilon}(y-y')\times\nonumber\\
&&(-\partial_{d}^{w}\delta(w-y))O_{y\,y'}
\oint_{\gamma} dz^{p}[f_{\rho}(y-z)
O_{z\,o\,y}+f_{\rho}(z-y)O_{z\,y}]\Delta_{cp}(\gamma_{o}^{z})]
\Psi(\gamma).
\end{eqnarray}

The following step is to integrate by parts
\begin{eqnarray}
&&\int d^{3}x \ut{N}(x)\int d^{3}w \ut{M}(w)\times\nonumber\\
&&\times\biggr(\oint_{\gamma} dy^{c}\oint_{\gamma} dy'^{d}
\delta(x-y')
f_{\epsilon}(y-y')\times\nonumber\\
&&(-\partial_{c}^{w}\delta(w-y))\,O_{y\,y'}
\oint_{\gamma} dz^{p}[f_{\rho}(y-z)
O_{z\,o\,y}+f_{\rho}(z-y)O_{z\,y}]\Delta_{dp}(\gamma_{o}^{z})]
\Psi(\gamma)-\nonumber\\
&&-\oint_{\gamma} dy^{c}\oint_{\gamma} dy'^{d} \delta(x-y')
f_{\epsilon}(y-y')\times\nonumber\\
&&(-\partial_{d}^{w}\delta(w-y))\,O_{y\,y'}
\oint_{\gamma} dz^{p}[f_{\rho}(y-z)
O_{z\,o\,y}+f_{\rho}(z-y)O_{z\,y}]\Delta_{cp}(\gamma_{o}^{z})]
\Psi(\gamma)\biggr)=\\
\nonumber\\
&&=-\int d^{3}x \ut{N}(x)\int d^{3}w\, D_{c}^{w}\biggr(\ut{M}(w)
\oint_{\gamma} dy^{c}\oint_{\gamma} dy'^{d} \delta(x-y')
f_{\epsilon}(y-y')\delta(w-y)
\times\nonumber\\
&&O_{y\,y'} \oint_{\gamma} dz^{p}\,f_{\rho}(y-z)\,[O_{z\,o\,y}+
O_{z\,y}]\Delta_{dp}(\gamma_{o}^{z})]\Psi(\gamma)\biggr)+\nonumber\\
&&+\int d^{3}x \ut{N}(x)\int d^{3}w\, \partial_{c}^{w}\ut{M}(w)
\times \oint_{\gamma} dy^{c}\oint_{\gamma} dy'^{d}
\delta(x-y') f_{\epsilon}(y-y')\delta(w-y)
\times\nonumber\\
&& O_{y\,y'} \oint_{\gamma} dz^{p}\,f_{\rho}(y-z)\,[O_{z\,o\,y}+
O_{z\,y}]\Delta_{dp}(\gamma_{o}^{z})\Psi(\gamma)+\nonumber\\
&&+\int d^{3}x \ut{N}(x)\int d^{3}w \,D_{d}^{w}\biggr(\ut{M}(w)
\oint_{\gamma} dy^{c}\oint_{\gamma} dy'^{d} \delta(x-y')
f_{\epsilon}(y-y')
\delta(w-y)\times\nonumber\\
&&O_{y\,y'} \oint_{\gamma} dz^{p}\,f_{\rho}(y-z)\,
[O_{z\,o\,y}+O_{z\,y}]\Delta_{cp}(\gamma_{o}^{z})]\Psi(\gamma)\biggr)
-\nonumber\\
&&-\int d^{3}x \ut{N}(x)\int d^{3}w \,\partial_{d}^{w}
\ut{M}(w)\times
\oint_{\gamma} dy^{c}\oint_{\gamma} dy'^{d} \delta(x-y')
f_{\epsilon}(y-y')\delta(w-y)\times\nonumber\\
&&O_{y\,y'} \oint_{\gamma} dz^{p}\,f_{\rho}(y-z)\,
[O_{z\,o\,y}+O_{z\,y}]\Delta_{cp}(\gamma_{o}^{z})]\Psi(\gamma),
\end{eqnarray}
and to notice that since we are considering only the case of compact
manifolds, the total derivatives that appear in the first and third
terms do not contribute. The resulting terms can be rewritten as,
\begin{eqnarray}
&&\int d^{3}x \int d^{3}w\,\ut{N}(x) \,\partial_{c}^{w}\ut{M}(w)
\times 2\oint_{\gamma} dy^{[c}\oint_{\gamma} dy'^{d]} \delta(x-y')
f_{\epsilon}(y-y')\delta(w-y)\times\nonumber\\
&&O_{y\,y'} \oint_{\gamma} dz^{p}\,f_{\rho}(y-z)\,
[O_{z\,o\,y}+O_{z\,y}]\Delta_{dp}(\gamma_{o}^{z})]\Psi(\gamma).
\end{eqnarray}

Taking into account the discussion at the beginning of this section,
it is very easy to see (due to the symmetry in changing $x$ and $w$
in the deltas) that the other term of the commutator corresponding
to the last one obtained is,
\begin{eqnarray}
&&-\int d^{3}x \int d^{3}w\,\ut{M}(w) \,\partial_{c}^{x}
\ut{N}(x)\times 2\oint_{\gamma} dy^{[c}\oint_{\gamma} dy'^{d]}
\delta(x-y')
f_{\epsilon}(y-y')\delta(w-y)\times\nonumber\\
&&O_{y\,y'} \oint_{\gamma} dz^{p}\,f_{\rho}(y-z)\,
[O_{z\,o\,y}+O_{z\,y}]\Delta_{dp}(\gamma_{o}^{z})]\Psi(\gamma).
\end{eqnarray}

The result then is,
\begin{eqnarray}
&&\int d^{3}x \int d^{3}w\,(\ut{N}(x) \,\partial_{c}^{w}\ut{M}(w)-
\ut{M}(w) \,\partial_{c}^{x}\ut{N}(x))\times\nonumber\\
&&\biggr(\oint_{\gamma} dy^{c}\oint_{\gamma} dy'^{d} \delta(x-y')
f_{\epsilon}(y-y') \delta(w-y)\times\nonumber\\
&&O_{y\,y'} \oint_{\gamma} dz^{p}\,f_{\rho}(y-z)\,
[O_{z\,o\,y}+O_{z\,y}]\Delta_{dp}(\gamma_{o}^{z})]\Psi(\gamma)-
\nonumber\\
&&-\oint_{\gamma} dy^{d} \oint_{\gamma} dy'^{c}  \delta(x-y')
f_{\epsilon}(y-y') \delta(w-y)\times\nonumber\\
&&O_{y\,y'} \oint_{\gamma} dz^{p}\,f_{\rho}(y-z)\,
[O_{z\,o\,y}+O_{z\,y}]\Delta_{dp}(\gamma_{o}^{z})]\Psi(\gamma)\biggr).
\end{eqnarray}

We now make some changes in this last expression so that we
can use the equality of appendix B. Noticing that,
\begin{eqnarray}
\delta(x-y') f_{\epsilon}(y-y') \delta(w-y)=
\delta(x-y') f_{\epsilon}(w-x) \delta(w-y)
\end{eqnarray}
we can rewrite the expression of interest as,
\begin{eqnarray}
&&\int d^{3}x \int d^{3}w\,(\ut{N}(x) \,\partial_{c}^{w}
\ut{M}(w)- \ut{M}(w) \,\partial_{c}^{x}\ut{N}(x))
f_{\epsilon}(w-x) \times\nonumber\\
&&\int d\overline{y}^{3} \delta(w-\overline{y})
\int d\overline{z}^{3} f_{\rho}(w-\overline{z})
\oint_{\gamma} dy^{c}\delta(y-\overline{y})\oint_{\gamma} dy'^{d}
\delta(y'-x)\times\nonumber\\
&&\biggr[O_{y\,y'} \oint_{\gamma} dz^{p}\delta(z-\overline{z})\,
[O_{z\,o\,y}+O_{z\,y}]\Delta_{dp}(\gamma_{o}^{z})]\Psi(\gamma)-
\nonumber\\
&&-O_{y'\,y} \oint_{\gamma} dz^{p}\delta(z-\overline{z})\,
[O_{z\,o\,y'}+O_{z\,y'}]\Delta_{dp}(\gamma_{o}^{z})]
\Psi(\gamma)\biggr],
\end{eqnarray}
and then as,
\begin{eqnarray}
&&\int d^{3}x \int d^{3}w\,(\ut{N}(x) \,\partial_{c}^{w}\ut{M}(w)-
\ut{M}(w) \,\partial_{c}^{x}\ut{N}(x)) f_{\epsilon}(w-x)
\times\nonumber\\
&&\int d\overline{y}^{3} \delta(w-\overline{y})
\int d\overline{z}^{3} f_{\rho}(w-\overline{z})
\oint_{\gamma} dy^{c}\delta(y-\overline{y})\oint_{\gamma} dy'^{d}
\delta(x-y')\times\nonumber\\
&&\oint_{O_{y\,y'}\gamma} dz^{p}\delta(z-\overline{z})\,[O_{y\,y'}
[O_{z\,o\,y}+O_{z\,y}]+O_{y'\,y}[O_{z\,o\,y'}+O_{z\,y'}]]
\Delta_{dp}(\gamma_{o}^{z})]\Psi(\gamma).\label{a9}
\end{eqnarray}

Finally, using the result of appendix B we can rewrite the above
expression in terms that only involve one rerouting operator, as
is required by the desired final result (which only involves one
rerouting coming from the metric that appears smearing the
diffeomorphism constraint). As a consequence of this the
commutator is,
\begin{eqnarray}
&&\lefteqn{\int d^{3}x \int d^{3}w\,(\ut{N}(x) \,\partial_{c}^{w}
\ut{M}(w)- \ut{M}(w) \,\partial_{c}^{x}\ut{N}(x))
f_{\epsilon}(w-x)  \times }\nonumber\\
&&\int d\overline{y}^{3} \delta(w-\overline{y})
\int d\overline{z}^{3} f_{\rho}(w-\overline{z})\times\nonumber\\
&&2\biggr[\oint_{\gamma} dy^{c}\delta(y-\overline{y})
\oint_{\gamma} dy'^{d}
\delta(x-y') \oint_{O_{y\,y'}\gamma} dz^{p}\delta(z-\overline{z})
\,[O_{z\,y}\Delta_{dp}(\gamma_{o}^{y'})-
O_{y'\,z}\Delta_{dp}(\gamma_{0}^{y})]\Psi(\gamma) \nonumber\\
&&+\oint_{\gamma} dy^{c}\delta(y-\overline{y})\oint_{\gamma} dy'^{d}
\delta(x-y')\times \oint_{\gamma} dz^{p}\delta(z-\overline{z})
[\frac{1}{2}\Delta_{dp}(\gamma_{o}^{y'})
-\frac{1}{2}\Delta_{dp}(\gamma_{o}^{y})]\Psi(\gamma)\biggr].\label{a7}
\end{eqnarray}

We now prove that in the above expression two of the terms cancel each
other and the other two give rise to the desired result.  Consider the
following two terms,
\begin{eqnarray}
&&\oint_{\gamma} dy^{c}\delta(y-\overline{y})\oint_{\gamma} dy'^{d}
\delta(x-y') \oint_{O_{y\,y'}\gamma} dz^{p}\delta(z-\overline{z})\,
O_{y'\,z}\Delta_{dp}(\gamma_{o}^{y})\Psi(\gamma) \nonumber\\
&&+\oint_{\gamma} dy^{c}\delta(y-\overline{y})\oint_{\gamma} dy'^{d}
\delta(x-y') \oint_{\gamma} dz^{p} \delta(z-\overline{z})\frac{1}{2}
\Delta_{dp}(\gamma_{o}^{y})\Psi(\gamma). \end{eqnarray}
We can rewrite
them as,
 \begin{eqnarray} &&\int
d^{3}\overline{y'}\delta(x-\overline{y'}) \oint_{\gamma}
dy^{c}\delta(y-\overline{y})\oint_{\gamma} dy'^{d}
\delta(\overline{y'}-y') \oint_{O_{y\,y'}\gamma}
dz^{p}\delta(z-\overline{z})\,
O_{y'\,z}\Delta_{dp}(\gamma_{o}^{y})\Psi(\gamma) \nonumber\\ &&+\int
d^{3}\overline{y'}\delta(x-\overline{y'}) \oint_{\gamma}
dy^{c}\delta(y-\overline{y})\oint_{\gamma} dy'^{d}
\delta(\overline{y'}-y') \oint_{\gamma} dz^{p}
\delta(z-\overline{z})\frac{1}{2}
\Delta_{dp}(\gamma_{o}^{y})\Psi(\gamma). \label{a5} \end{eqnarray}

It is convenient to write the product of the last two loop integrals
symbolically and study them with diagrams, assuming for example that
$y'$ precedes $y$ in the loop $\gamma$.  The alternative case can be
studied in a similar way.  So,
\begin{eqnarray} &&\int_{\gamma_{o\,y}} dy'^{d}
\biggr[\ \int_{\gamma_{o\,y'}} dz^{p}\ + \
\int_{\overline{\gamma}_{y\,y'}} dz^{p}+ \ \int_{\gamma_{y\,o}} dz^{p}
\biggr]\ + \end{eqnarray} \unitlength=0.50mm

\linethickness{0.4pt}
\begin{picture}(216.92,24.11)
\put(103.08,18.46){\makebox(0,0)[cc]{o}}
\put(110.77,18.46){\makebox(0,0)[cc]{z}}
\put(118.97,18.46){\makebox(0,0)[cc]{y'}}
\put(127.18,17.43){\makebox(0,0)[cc]{y}}
\put(148.21,18.46){\makebox(0,0)[cc]{o}}
\put(155.90,18.46){\makebox(0,0)[cc]{y'}}
\put(164.10,18.46){\makebox(0,0)[cc]{z}}
\put(171.79,17.43){\makebox(0,0)[cc]{y}}
\put(189.74,18.46){\makebox(0,0)[cc]{z}}
\put(197.95,18.46){\makebox(0,0)[cc]{o}}
\put(206.15,18.46){\makebox(0,0)[cc]{y'}}
\put(213.85,17.43){\makebox(0,0)[cc]{y}}
\put(103.08,23.08){\line(0,1){1.03}}
\put(103.08,24.10){\line(0,-1){2.05}}
\put(103.08,22.05){\line(0,1){1.03}}
\put(103.08,23.08){\line(1,0){7.69}}
\put(110.77,23.08){\line(0,1){1.03}}
\put(110.77,24.10){\line(0,-1){2.05}}
\put(110.77,22.05){\line(0,1){1.03}}
\put(110.77,23.08){\line(1,0){8.21}}
\put(118.97,23.08){\line(0,1){1.03}}
\put(118.97,24.10){\line(0,-1){2.05}}
\put(118.97,22.05){\line(0,1){1.03}}
\put(118.97,23.08){\line(1,0){8.21}}
\put(127.18,23.08){\line(0,1){1.03}}
\put(127.18,24.10){\line(0,-1){2.05}}
\put(127.18,22.05){\line(0,1){1.03}}
\put(148.21,23.08){\line(0,1){1.03}}
\put(148.21,24.10){\line(0,-1){2.05}}
\put(148.21,22.05){\line(0,1){1.03}}
\put(148.21,23.08){\line(1,0){7.69}}
\put(155.90,23.08){\line(0,1){1.03}}
\put(155.90,24.10){\line(0,-1){2.05}}
\put(155.90,22.05){\line(0,1){1.03}}
\put(155.90,23.08){\line(1,0){8.21}}
\put(164.10,23.08){\line(0,1){1.03}}
\put(164.10,24.10){\line(0,-1){2.05}}
\put(164.10,22.05){\line(0,1){1.03}}
\put(164.10,23.08){\line(1,0){7.69}}
\put(171.79,23.08){\line(0,1){1.03}}
\put(171.79,24.10){\line(0,-1){2.05}}
\put(171.79,22.05){\line(0,1){1.03}}
\put(189.74,23.08){\line(0,1){1.03}}
\put(189.74,24.10){\line(0,-1){2.05}}
\put(189.74,22.05){\line(0,1){1.03}}
\put(189.74,23.08){\line(1,0){8.21}}
\put(197.95,23.08){\line(0,1){1.03}}
\put(197.95,24.10){\line(0,-1){2.05}}
\put(197.95,22.05){\line(0,1){1.03}}
\put(197.95,23.08){\line(1,0){8.21}}
\put(206.15,23.08){\line(0,1){1.03}}
\put(206.15,24.10){\line(0,-1){2.05}}
\put(206.15,22.05){\line(0,1){1.03}}
\put(206.15,23.08){\line(1,0){7.69}}
\put(213.85,23.08){\line(0,1){1.03}}
\put(213.85,24.10){\line(0,-1){2.05}}
\put(213.85,22.05){\line(0,1){1.03}}
\put(100.00,23.08){\line(1,0){29.74}}
\put(145.13,23.08){\line(1,0){29.74}}
\put(187.18,23.08){\line(1,0){29.74}}
\end{picture}

\begin{eqnarray}
&&+\int_{\gamma_{y\,o}}  dy'^{d} \biggr[
\ \int_{\overline{\gamma}_{y\,o}}  dz^{p}\ +
\ \int_{\gamma_{y\,y'}} dz^{p}\ +
\ \int_{\overline{\gamma}_{o\,y'}}  dz^{p} \biggr]=
\end{eqnarray}
\unitlength=0.50mm
\linethickness{0.4pt}
\begin{picture}(222.05,14.88)
\put(107.18,12.82){\line(0,1){2.05}}
\put(107.18,14.87){\line(0,-1){1.03}}
\put(107.18,13.85){\line(1,0){7.69}}
\put(114.87,13.85){\line(0,1){1.03}}
\put(114.87,14.87){\line(0,-1){2.05}}
\put(114.87,12.82){\line(0,1){1.03}}
\put(114.87,13.85){\line(1,0){8.21}}
\put(123.08,13.85){\line(0,1){1.03}}
\put(123.08,14.87){\line(0,-1){2.05}}
\put(123.08,12.82){\line(0,1){1.03}}
\put(123.08,13.85){\line(1,0){7.69}}
\put(130.77,13.85){\line(0,1){1.03}}
\put(130.77,14.87){\line(0,-1){2.05}}
\put(130.77,12.82){\line(0,1){1.03}}
\put(151.79,13.85){\line(0,1){1.03}}
\put(151.79,14.87){\line(0,-1){2.05}}
\put(151.79,12.82){\line(0,1){1.03}}
\put(151.79,13.85){\line(1,0){8.21}}
\put(160.00,13.85){\line(0,1){1.03}}
\put(160.00,14.87){\line(0,-1){2.05}}
\put(160.00,12.82){\line(0,1){1.03}}
\put(160.00,13.85){\line(1,0){8.21}}
\put(168.21,13.85){\line(0,1){1.03}}
\put(168.21,14.87){\line(0,-1){2.05}}
\put(168.21,12.82){\line(0,1){1.03}}
\put(168.21,13.85){\line(1,0){7.69}}
\put(175.90,13.85){\line(0,1){1.03}}
\put(175.90,14.87){\line(0,-1){2.05}}
\put(175.90,12.82){\line(0,1){1.03}}
\put(194.87,13.85){\line(0,1){1.03}}
\put(194.87,14.87){\line(0,-1){2.05}}
\put(194.87,12.82){\line(0,1){1.03}}
\put(194.87,13.85){\line(1,0){8.21}}
\put(203.08,13.85){\line(0,1){1.03}}
\put(203.08,14.87){\line(0,-1){2.05}}
\put(203.08,12.82){\line(0,1){1.03}}
\put(203.08,13.85){\line(1,0){7.69}}
\put(210.77,13.85){\line(0,1){1.03}}
\put(210.77,14.87){\line(0,-1){2.05}}
\put(210.77,12.82){\line(0,1){1.03}}
\put(210.77,13.85){\line(1,0){8.21}}
\put(218.97,13.85){\line(0,1){1.03}}
\put(218.97,14.87){\line(0,-1){2.05}}
\put(218.97,12.82){\line(0,1){1.03}}
\put(194.87,13.85){\line(-1,0){3.08}}
\put(218.97,13.85){\line(1,0){3.08}}
\put(175.90,13.85){\line(1,0){3.08}}
\put(151.79,13.85){\line(-1,0){2.56}}
\put(130.77,13.85){\line(1,0){3.08}}
\put(107.18,13.85){\line(-1,0){3.08}}
\put(107.18,9.23){\makebox(0,0)[cc]{z}}
\put(114.87,8.20){\makebox(0,0)[cc]{y}}
\put(123.08,9.23){\makebox(0,0)[cc]{y'}}
\put(130.77,9.23){\makebox(0,0)[cc]{o}}
\put(151.79,8.20){\makebox(0,0)[cc]{y}}
\put(160.00,9.23){\makebox(0,0)[cc]{z}}
\put(168.21,9.23){\makebox(0,0)[cc]{y'}}
\put(175.90,9.23){\makebox(0,0)[cc]{o}}
\put(194.87,8.20){\makebox(0,0)[cc]{y}}
\put(203.08,9.23){\makebox(0,0)[cc]{y'}}
\put(210.77,9.23){\makebox(0,0)[cc]{z}}
\put(218.97,9.23){\makebox(0,0)[cc]{o}}
\end{picture}
\begin{eqnarray}
&&=\int_{\gamma_{o\,y}}  dz^{p}
\biggr[\ \int_{\gamma_{z\,y}}  dy'^{d}\ +
\ \int_{\overline{\gamma}_{z\,o}}  dy'^{d}+
\ \int_{\overline{\gamma}_{o\,y}}  dy'^{d} \biggr]\ +\nonumber\\
&&+\int_{\gamma_{y\,o}}  dz^{p} \biggr[
\ \int_{\overline{\gamma}_{z\,y}}  dy'^{d}\ +
\ \int_{\gamma_{z\,o}} dy'^{d}\ +
\ \int_{\gamma_{o\,y}}  dy'^{d} \biggr].
\end{eqnarray}

We can then rewrite  expression (\ref{a5}) as,
\begin{eqnarray}
&&\int d^{3}\overline{y'}\delta(x-\overline{y'})
\oint_{\gamma} dy^{c}\delta(y-\overline{y})\oint_{\gamma} dz^{p}
\delta(\overline{z}-z)
\oint_{O_{z\,y}\gamma} dy'^{d}\delta(y'-\overline{y'})\,
O_{y'\,z}\Delta_{dp}(\gamma_{o}^{y})\Psi(\gamma) \nonumber\\
&&+\int d^{3}\overline{y'}\delta(x-\overline{y'})
\oint_{\gamma} dy^{c}\delta(y-\overline{y})\oint_{\gamma} dz^{p}
\delta(\overline{z}-z)
\oint_{\gamma} dy'^{d} \delta(y'-\overline{y'})\frac{1}{2}
\Delta_{dp}(\gamma_{o}^{y})\Psi(\gamma),
\end{eqnarray}
and  making the change $y'$ by $z$ (and removing both
regulators so that one can change $ \overline{y'}$ by $
\overline{z}$ only in the last two deltas) we get,
\begin{eqnarray}
&&\int d^{3}\overline{y'}\delta(x-\overline{y'})
\oint_{\gamma} dy^{c}\delta(y-\overline{y})\oint_{\gamma} dy'^{p}
\delta(\overline{y'}-y')
\oint_{O_{z\,y}\gamma} dz^{d}\delta(z-\overline{z})\,
O_{z\,y'}\Delta_{dp}(\gamma_{o}^{y})\Psi(\gamma) \nonumber\\
&&+\int d^{3}\overline{y'}\delta(x-\overline{y'})
\oint_{\gamma} dy^{c}\delta(y-\overline{y})\oint_{\gamma} dy'^{p}
\delta(\overline{y'}-y')
\oint_{\gamma} dz^{d} \delta(z-\overline{z})\frac{1}{2}
\Delta_{dp}(\gamma_{o}^{y})\Psi(\gamma).
\end{eqnarray}

Finally, taking into account that $ O_{y'\,z}
\Delta_{dp}(\gamma_{o}^{y})=
- O_{z\,y'} \Delta_{dp}(\gamma_{o}^{y})$ , and changing $d$ by $p$ we
get,
\begin{eqnarray}
&&- \int d^{3}\overline{y'}\delta(x-\overline{y'})
\oint_{\gamma} dy^{c}\delta(y-\overline{y})\oint_{\gamma} dy'^{d}
\delta(\overline{y'}-y')
\oint_{O_{y\,y'}\gamma} dz^{p}\delta(z-\overline{z})\,
O_{y'\,z}\Delta_{dp}(\gamma_{o}^{y})\Psi(\gamma) \nonumber\\
&&-\int d^{3}\overline{y'}\delta(x-\overline{y'})
\oint_{\gamma} dy^{c}\delta(y-\overline{y})\oint_{\gamma} dy'^{d}
\delta(\overline{y'}-y') \oint_{\gamma} dz^{p}
\delta(z-\overline{z})\frac{1}{2}\Delta_{dp}(\gamma_{o}^{y})
\Psi(\gamma).
\end{eqnarray}

We have therefore proved that expression \ref{a5} is equal to minus
itself.  Using the same tools one can demonstrate that the two
remaining terms in \ref{a7} are equivalent to,
\begin{eqnarray}
&&\lefteqn{\int d^{3}x \int d^{3}w\,(\ut{N}(x) \,\partial_{c}^{w}
\ut{M}(w)-
\ut{M}(w) \,\partial_{c}^{x}\ut{N}(x)) f_{\epsilon}(w-x)
\times } \nonumber\\
&&\int d\overline{y}^{3} \delta(w-\overline{y})
\int d\overline{z}^{3} f_{\rho}(w-\overline{z}) \times\nonumber\\
&&2\biggr[\oint_{\gamma} dy^{c}\delta(y-\overline{y})
\oint_{\gamma} dy'^{d}
\delta(x-y') \oint_{O_{y\,y'}\gamma} dz^{p}\delta(z-\overline{z})
\,O_{y\,y'}\Delta_{dp}(\gamma_{o}^{z})
\Psi(\gamma) \hspace{3cm}\nonumber\\
&&+\oint_{\gamma} dy^{c}\delta(y-\overline{y})\oint_{\gamma} dy'^{d}
\delta(x-y')\times \oint_{\gamma} dz^{p}\delta(z-\overline{z})
\frac{1}{2}\Delta_{dp}(\gamma_{o}^{z})\Psi(\gamma)\biggr].
\end{eqnarray}

Inserting the factor $\frac {\epsilon}{4}$ that was omitted, this last
expression is the regularized form of the product of the metric
operator times the diffeomorphism constraint, which is consistent with
the classical result (see appendix C).

\section{Conclusions}

In this paper we have studied the formal commutation relations of the
constraints of quantum gravity in the loop representation. We have
shown that the expressions available for the constraint are
operationally useful to compute the commutators. We use a background
dependent regulator and only recover the correct commutation relations
in the formal limit in which the regulator is removed. Our
computations set the stage for a future computation keeping the
regulators at the lowest order.

\section{Acknowledgements}

This work was supported in part by grant NSF-PHY93-96246,
NSF-PHY-92-07225
and by research funds from the University of
Utah and The Pennsylvania State University. We also acknowledge
support of Pedeciba and Conicyt (Uruguay).

\appendix
\section{}

We need a gauge invariant regularization prescription. If one
performs a point-splitting at intersections, one needs to close
the loops to preserve gauge invariance after the rerouting process.
In the following example we denote the intersecting point as $I$.
We will use the following notation in this appendix:
$\gamma_{y\,y'}$ denotes the portion of $\gamma$ from $y$ to $y'$
without passing through the origin no matter the original orientation
of that portion.  The other cases that appear have an analogous meaning.
Explicitly, the gauge invariance-preserving action of the rerouting
operator can be defined as,
\begin{eqnarray}
f_{\epsilon}(y-y')\ O_{y\,y'}\ \Psi(\gamma)=
f_{\epsilon}(y-y')\Psi(O_{y\,y'}\gamma)=\nonumber\\
=f_{\epsilon}(y-y')\ \Psi(\gamma_{y\,y'}
\circ\gamma_{y'\,I}\circ
\gamma_{I\,y} \circ\gamma_{y\,o\,y'} \circ \gamma_{y'\,I}
\circ \gamma_{I\,y}).
\end{eqnarray}

Let us now consider the action of the rerouting operator on the loop
derivative.  Consider that the point $z$ is in the piece
$\gamma_{y'\,o\,y}$ ($y$ before $y'$, for example) \cite{foot2}. Then,
\begin{eqnarray}
&&O_{y\,y'}\ \Delta_{d\,p}(\gamma_{o}^{z}) \Psi(\gamma)=
-\Delta_{d\,p}(\gamma_{o}^{z})O_{y\,y'} \Psi(\gamma)=
\Delta_{d\,p}([O_{y\,y'}\gamma]_{o}^{z}) \Psi(O_{y\,y'}\gamma).
\end{eqnarray}

The minus sign appears because the rerouting affects the loop
dependence of the loop derivative. It is a situation analogous to
action of the parity operator in quantum mechanics on the partial
derivative,
\begin{eqnarray}
P\partial_{x}\Psi(x)=- \partial_{x} P\Psi(x)=\partial_{x'}\Psi(x').
\end{eqnarray}
If the point $z$ of the loop derivative is in $\gamma_{y\,y'}$ there is
no sign change.

Let us now consider specifically the case that appears in the
Hamiltonian constraint, where the loop derivative is evaluated at one
of the points of the action of the rerouting operator,
\begin{eqnarray}
\sigma^{d\,p}\ f_{\epsilon}(y-y')\ O_{y\,y'}
\ \Delta_{d\,p}(\gamma_{o}^{y}) \Psi(\gamma)\equiv
f_{\epsilon}(y-y')\ \delta_{y} \Psi(\gamma_{y\,y'} \circ \gamma_{y'\,I}
\circ \gamma_{I\,y}\circ \gamma_{y\,o\,y'} \circ \gamma_{y'\,I}
\circ \gamma_{I\,y})
\end{eqnarray}
where the result of the operator $\delta_y$
\begin{eqnarray}
\delta_{y} \Psi(\gamma_{y\,y'} \circ \gamma_{y'\,I}
\circ \gamma_{I\,y}\circ \gamma_{y\,o\,y'} \circ \gamma_{y'\,I}
\circ \gamma_{I\,y}) = &&\Psi(\delta \gamma \circ
\gamma_{y\,y'} \circ \gamma_{y'\,I}
\circ \gamma_{I\,y}\circ \gamma_{y\,o\,y'} \circ \gamma_{y'\,I}
\circ \gamma_{I\,y})\label{app5}\\
&&-
\Psi(\gamma_{y\,y'} \circ \gamma_{y'\,I}
\circ \gamma_{I\,y}\circ \gamma_{y\,o\,y'} \circ \gamma_{y'\,I}
\circ \gamma_{I\,y})\nonumber
\end{eqnarray}

Notice that the definition of the operator $\delta_y$ implies that the
action of the loop derivative is with respect to $\gamma$ and the
rerouting operator acts {\em afterwards}.

\section{}

In this appendix we will prove that expressions (\ref{a9}) and
(\ref{a7}),
are equal \cite{foot2}. First we write the Mandelstam identities for
wavefunctions. For loops $ \gamma, \gamma_{1}, \gamma_{2}$, and
$\gamma_{3}$,
\begin{eqnarray} &&\Psi(\gamma)=\Psi(\gamma^{-1})\\
&&\Psi(\gamma_{1}\circ\gamma_{2})=\Psi(\gamma_{2}\circ\gamma_{1})\\
&&\Psi(\gamma_{1}\circ\gamma_{2}\circ\gamma_{3})+\
\Psi(\gamma_{1}\circ\gamma_{2}\circ\gamma_{3}^{-1})=
\Psi(\gamma_{2}\circ\gamma_{1}\circ\gamma_{3})+
\Psi(\gamma_{2}\circ\gamma_{1}\circ\gamma_{3}^{-1}).
\end{eqnarray}

Considering equation (\ref{app5}), (with $p=z$) we can write the
action of the loop operators in expression (\ref{a9}),
\begin{eqnarray}
\delta_{z}\Psi(O_{y\,y'}O_{z\,o\,y}\gamma)+\delta_{z}
\Psi(O_{y\,y'}O_{z\,y}\gamma)+ \delta_{z}\Psi(O_{y'\,y}O_{z\,o\,y'}
\gamma)+\delta_{z}\Psi(O_{y'\,y}O_{z\,y'}\gamma)
\end{eqnarray}

Additionally, using the results of the appendixes A and the Mandelstam
identities of Appendix B we can show\\
(p is the intersecting point of $\gamma_{1}\, \gamma_{2}\,
\gamma_{3}$),
\begin{eqnarray}
&&\delta_{p}\Psi(\gamma_{1}\circ\overline{\gamma}_{2}\circ\gamma_{3})+
\delta_{p}\Psi(\gamma_{3}\circ\gamma_{1}\circ\overline{\gamma}_{2})+
\delta_{p}\Psi(\overline{\gamma}_{2}\circ\overline{\gamma}_{1}
\circ\gamma_{3})+
\delta_{p}\Psi(\overline{\gamma}_{1
}\circ\gamma_{3}\circ\overline{\gamma}_{2})=\nonumber\\
&&2[\delta_{p}\Psi(\gamma_{3}\circ\gamma_{2}
\circ\overline{\gamma}_{1})+
\frac{1}{2} \delta_{p}\Psi(\gamma_{3}\circ\gamma_{2}\circ\gamma_{1})-
\delta_{p}\Psi(\gamma_{1}\circ\overline{\gamma}_{3}\circ\gamma_{2})-
\frac{1}{2} \delta_{p}\Psi(\gamma_{1}\circ\gamma_{3}\circ\gamma_{2})].
\end{eqnarray}

Using this last identity, the integrand in (\ref{a9}) can be written
as,
\begin{eqnarray}
2[\delta_{y'}\Psi(O_{z\,y}\gamma)+\frac{1}{2}\delta_{y'}\Psi(\gamma)
-\delta_{y}\Psi(O_{y'\,z}\gamma)-\frac{1}{2}\delta_{y}\Psi(\gamma)].
\end{eqnarray}

On the other hand, we can make a partition in the loop integrals
so that expression (\ref{a9}) splits in 6 terms. Symbolically we can
write,
\begin{eqnarray}
&&\oint_{\gamma} dy^{c} \oint_{\gamma} dy'^{d} \oint_{O_{y\,y'}
\gamma} dz^{p}=
\oint_{\gamma} dy^{c}\biggr[ \oint_{\gamma_{o\,y}} dy'^{d}+
\oint_{\gamma_{y\,o}} dy'^{d}\biggr]\times\nonumber\\
&&\times\biggr[\oint_{O\gamma_{y\,o}} dz^{p}+
\oint_{O\gamma_{o\,y'}} dz^{p}+\oint_{O\gamma_{y\,y'}} dz^{p}\biggr].
\end{eqnarray}
Then, combining (B6) and (B7), we finally get expression
(\ref{a7}), which in shorthand writing is,
\begin{eqnarray}
&&2\oint_{\gamma} dy^{c} \oint_{\gamma} dy'^{d}\oint_{O_{y\,y'}
\gamma}dz^{p}
[O_{z\,y}\Delta_{d\,p}(\gamma_{o}^{y'})-O_{y'\,z}
\Delta_{d\,p}(\gamma_{o}^{y})]+\nonumber\\
&&+2\oint_{\gamma} dy^{c} \oint_{\gamma} dy'^{d}\oint_{\gamma}dz^{p}
[\frac{1}{2}\Delta_{d\,p}(\gamma_{o}^{y'})-\frac{1}{2}
\Delta_{d\,p}(\gamma_{o}^{y})]
\end{eqnarray}

As an aid to visualize the meaning of the identity (B5), its
counterpart in the connection representation is the following identity
between $SU(2)$ matrices,
\begin{eqnarray}
\lefteqn{Tr[x^{d}BA^{-1}C]+Tr[x^{d}CBA^{-1}]+
Tr[x^{d}A^{-1}B^{-1}C]+Tr[x^{d}B^{-1}CA^{-1}]=}  \nonumber\\
&&2[Tr[Bx^{d}C]\  Tr[A]-\frac{1}{2}Tr[ABx^{d}C]-
Tr[Ax^{D}B]\  Tr[C]+\frac{1}{2}Tr[Ax^{D}BC]] \nonumber\\
&&(=2[\frac{1}{2}Tr[ABx^{d}C]+Tr[Bx^{d}CA^{-1}]-
\frac{1}{2}Tr[Ax^{D}BC]-Tr[Ax^{D}BC^{-1}]]).
\end{eqnarray}
Where $x^{d}=\frac{1}{2}\sigma^{d}$,\  and $\sigma^{d}
\  1\leq{d}\leq{3}$ are the Pauli matrices and $A,B$  and $C$,
are complex $SU(2)$ matrices.
This identity can be easily proved noting that,
\begin{equation}
Tr[x^{d}PQ^{-1}]+Tr[x^{d}Q^{-1}P]=2\,Tr[x^{d}P]\,Tr[Q]-
Tr[x^{d}PQ]-Tr[x^{d}QP]
\end{equation}
and,
\begin{eqnarray}
&&Tr[P^{-1}x^{d}Q]=-Tr[x^{d}P]\, Tr[Q]+Tr[x^{d}PQ]\\
&&Tr[Qx^{d}P^{-1}]=Tr[x^{d}QP]-Tr[x^{d}P]\,Tr[Q].
\end{eqnarray}

\section{}

To obtain the expression for the metric operator in the loop
representation we first compute it in the connection representation
and apply the loop version. The action of the metric operator is,
\begin{eqnarray}
&&f_{\epsilon}(\overline{w}-\overline{z})
\frac {\delta} {\delta A(\overline{w})_{b} ^{k}}
\frac {\delta} {\delta A(\overline{z})_{a} ^{k}} Tr[U(\gamma)]=
f_{\epsilon}(\overline{w}-\overline{z})\oint_{\gamma} dz^{a}
\delta(z-\overline{z})\times\nonumber\\
&&\frac {1}{2} \biggr[\oint_{\gamma_{o\,z}} dw^{b}
\delta(w-\overline{w}) Tr[U(\gamma_{w\,z})
U(\overline{\gamma}_{w\,o\,z}]+
\oint_{\gamma_{z\,o}} dw^{b} \delta(w-\overline{w}) Tr[U(\gamma_{z\,w})
U(\overline{\gamma}_{z\,o\,w}]\biggr]+\nonumber\\
&&+f_{\epsilon}(\overline{w}-\overline{z})\frac{1} {4}
\oint_{\gamma} dz^{a} \delta(z-\overline{z})
\oint_{\gamma} dw^{b} \delta(w-\overline{w}) Tr[U(\gamma)]=\nonumber\\
&&=\frac {1}{2} f_{\epsilon}(\overline{w}-\overline{z})
\oint_{\gamma} dz^{a} \delta(z-\overline{z})\biggr[
\oint_{\gamma_{o\,z}} dw^{b} \delta(w-\overline{w}) O_{w\,z}
Tr[U(\gamma)]+
\oint_{\gamma_{z\,o}} dw^{b} \delta(w-\overline{w}) O_{z\,w}
Tr[U(\gamma)]
\biggr]\nonumber\\
&&+f_{\epsilon}(\overline{w}-\overline{z})\frac{1} {4}
\oint_{\gamma} dz^{a} \delta(z-\overline{z})
\oint_{\gamma} dw^{b} \delta(w-\overline{w}) Tr[U(\gamma)]=\nonumber\\
&&=\frac {1}{2}f_{\epsilon}(\overline{w}-\overline{z})
\oint_{\gamma} dz^{a} \delta(z-\overline{z})
\oint_{\gamma} dw^{b} \delta(w-\overline{w}) O_{z\,w} Tr[U(\gamma)]
+\nonumber\\
&&+f_{\epsilon}(\overline{w}-\overline{z})\frac{1} {4}
\oint_{\gamma} dz^{a} \delta(z-\overline{z})
\oint_{\gamma} dw^{b} \delta(w-\overline{w}) Tr[U(\gamma)]
\end{eqnarray}
Where we have used that $O_{w\,z} Tr[U(\gamma)]=O_{z\,w}
Tr[U(\gamma)]$. Recall, when comparing with section VI, that we have
omitted a factor ${1\over 4}$ in the commutator of the Hamiltonians.

\end{document}